\documentstyle[preprint,revtex,eqsecnum]{aps}
\begin{document}
\draft
\begin{title}
The sensitivity of the Laser Interferometer Gravitational\\
Wave Observatory (LIGO) to a stochastic background,\\
and its dependence on the detector orientations.
\end{title}
\author{Eanna E. Flanagan}
\begin{instit}
Theoretical Astrophysics, California Institute of Technology,
Pasadena, California 91125
\end{instit}
\begin{abstract}

We analyze the sensitivity of a network of interferometer
gravitational-wave detectors to the gravitational-wave stochastic
background, and derive the dependence of this sensitivity on the
orientations of the detector arms.  We build on and extend the recent
work of Christensen, but our conclusion for the optimal choice of
orientations of a pair of detectors differs from his.  For a pair of
detectors (such as LIGO) that subtends an angle at the center of the
earth of $\,\alt 70^\circ$, we find that the optimal configuration is
for each detector to have its arms make an angle of $45^\circ$ (modulo
$90^\circ$) with the arc of the great circle that joins them.  For
detectors that are farther separated, each detector should instead
have one arm aligned with this arc.  We show that the broadband
sensitivity to the stochastic background of a detector pair which are
$\alt 3000 \, {\rm km}$ apart is essentially determined by their
relative rotation.  Their average rotation with respect to the arc
joining them is unimportant.  We also describe in detail the optimal
data-analysis algorithm for searching for the stochastic background
with a detector network, which is implicit in earlier work of
Michelson.  The LIGO pair of detectors will be separated by $\sim 3000
\, {\rm km}$.  The minimum detectable stochastic energy-density for
these detectors with their currently planned orientations is $\sim
3\%$ greater than what it would be if the orientations were optimal,
and $\sim 4$ times what it would be if their separation were $\alt$ a
few kilometers.  [The detectors are chosen to be far apart so that
their sources of noise will be uncorrelated, and in order to improve
the angular resolution of the determinations of positions of burst
sources.]

\end{abstract}
\newpage

\section{INTRODUCTION AND SUMMARY}
\label{intro}

\subsection{Background and motivation}

Construction will begin soon on the American Laser Interferometer
Gravitational Wave Observatory (LIGO) \cite{ligo_science}, and on its
French/Italian counterpart, VIRGO \cite{virgo}.  Early in the next
century there will likely be in operation a worldwide network of
detectors, with sites in America, Europe, and possibly Japan and
Australia \cite{Blair}.  It is important at this stage for physicists
to look ahead and identify the types of science that the community
might focus on using this network when it reaches a mature stage,
perhaps a decade after the first gravitational waves are detected.
One of the reasons for doing so is that some properties and parameters
of the network, which ultimately will constrain what it can accomplish
in the future, are being finalized today.  The orientation of the
detector arms is one example.

One of the long term aims of this detector network will be to place
upper limits on (or perhaps detect) the energy density of a stochastic
background (SB) of gravitational waves.  This background would be
analogous to the relic $3 {}^\circ {\rm K}$ electromagnetic
background, except that its spectrum is not expected to be thermal.
The spectrum is usually characterized by a quantity $\Omega_g(f)$
which is the gravitational-wave energy density per unit logarithmic
frequency, divided by the critical energy density $\rho_c$ to close
the Universe:
\begin{equation}
\label{omega_def}
\Omega_g(f) = {1
\over \rho_{\rm c} } {d E \over d^3x \, d (\ln f) }.
\end{equation}
Some possible sources of a SB include: (i) random superposition of
many weak signals from binary-star systems \cite{binaries}, (ii)
decaying cosmic strings\cite{strings} and first order phase
transitions\cite{phase} in the early Universe, and (iii) parametric
amplification of quantum mechanical zero-point fluctuations in the
metric tensor during inflation \cite{relicgs,Wise,White}.  See
Refs.~\cite{300years,prospect,thesis} for reviews.  The predicted wave
strengths from all of these stochastic sources are highly uncertain,
reflecting our relative ignorance of the relevant physics and/or
astrophysics.  Hence, detecting or placing upper limits on the SB can
bring us valuable information, particularly about the very early
Universe.

Relic gravitational waves produced during inflation are particularly
interesting, because, as Grischuk has shown \cite{imprint}, the energy
spectrum for these waves contains a unique imprint of the time
evolution of the Universe's scale factor $a(t)$.  We now discuss what
is known about the magnitude of the contribution to $\Omega_g(f)$ from
these waves, at frequencies relevant to LIGO/VIRGO.  The predictions
for $\Omega_g$ from cosmological models are not very firm: they can
vary between $\sim 1$ and $\sim 10^{-14}$ or less.  However,
observational upper bounds on $\Omega_g(f)$ in various frequency bands
give interesting constraints on the models \cite{Wise}.  In turn these
constraints can be used, within the context of particular inflation
models, to place upper bounds on the contribution of relic gravitons
to the value of $\Omega_g(f)$ in the LIGO/VIRGO waveband, see, e.g.,
Ref.~\cite{Starobinsky}.  The strongest such upper bound comes from
matching the normalization of the scalar and tensor fluctuations
produced during inflation to the recent COBE measurement of the
microwave background anisotropy \cite{White,Davies}.  The result of
this matching is somewhat discouraging: standard exponential inflation
models predict that $\Omega_g(f) \le 3 \times 10^{-14}$ at the $95\%$
confidence level \cite{White}, far too small to be detected
[cf.~Eq.~$(\ref{ultimate})$ below].  While it is far from certain that
exponential inflation is correct, it seems unlikely that the expansion
during the inflationary era was so much faster than exponential as to
tilt the gravitational wave spectrum enough to give a detectable
signal at high, LIGO/VIRGO frequencies.  It is also possible, of
course, that the observed microwave anisotropy was caused by physical
processes of structure formation other than inflation.

Despite these pessimistic prospects for the detection of relic
gravitons by LIGO/VIRGO, it is certainly possible that there will be a
detectable signal from other sources such as cosmic strings
\cite{strings}.  Hence, it is important to determine how the detector
arm orientations, which will not be changeable in the future, affect
the sensitivity of the detector network to the SB.  To do so is the
first of the two principal purposes of this paper.  This issue was
first considered by Michelson \cite{mi}, and has been extensively
discussed by Christensen \cite{thesis,paper}.  Essentially we build on
and extend slightly their analyses.  Our conclusions are also slightly
different from those of Christensen.

The second principal purpose of this paper is to spell out the optimal
data-processing procedure for searching for the SB with a
network of detectors.  The algorithm for two detectors is implicit in
Michelson \cite{mi} (and is incorrectly treated in Ref.~\cite{paper});
we give a more detailed description and a generalization to a network
of detectors, taking into account the possible effect of correlated
sources of noise.  We now turn to a description of our results and an
explanation of how they relate to earlier work.

\subsection{Detection of the stochastic background}

The effect of the SB on a gravitational-wave detector is essentially
to produce a small contribution to the random, gaussian noise in its
output.  For one detector this contribution will be swamped by the
detector's own sources of noise, unless the SB strength is implausibly
large ($\Omega_g \sim 10^{-6}$; see Sec.~\ref{correlations}).  For two
detectors which have no common sources of noise, however, the only
contribution to the correlated fluctuations in the detector outputs
will be the SB.  By cross correlating the outputs of the detectors,
the SB can in principle be measured.  If one had identical, LIGO type
detectors at the same site, oriented in same way so that they respond
in exactly the same way to the SB, and with levels of intrinsic noise
corresponding to the ``advanced detectors'' of
Ref.~\cite{ligo_science}, then cross correlating would give a
sensitivity to $\Omega_g$ of the order of $10^{-10}$ in the frequency
band $10\,{\rm Hz} \alt f \alt 1000\,{\rm Hz}$ \cite{300years}.

For a pair of separated, non-aligned detectors, two new physical
effects complicate the analysis \cite{thesis,mi,paper}.  First, if the
detectors are not aligned the same way, they will respond to different
polarization components of the SB.  Orthogonal polarization components
of the SB are expected to be statistically independent, and so the
cross correlation will be reduced.  Second, for each mode there will
be a time lag between exciting the first detector and the second
detector, and hence phase lags in the cross correlation.  For the LIGO
detectors separated by $\sim 3000 \, {\rm km}$ and having maximum
sensitivity at a frequency of $f \sim 70 \, {\rm Hz}$, a typical phase
lag will be of order unity.  Hence, there will be some destructive
interference between parts of the cross-correlation that are due to
modes which propagate in different directions.  Thus we expect a
reduction in the sensitivity of the detector pair to the SB.

To analyze this reduction, it is necessary to (i) determine how to
{\it optimally process} the output from the detectors, and (ii) find
the signal to noise ratio (SNR) that results from this method of
filtering.  We also want to (iii) determine how the optimal SNR
depends on the detector orientations, and (iv) find those orientations
that maximize the SNR.  Steps (i) and (ii) were analyzed in
Refs.~\cite{mi,paper}.  They found that, when optimal signal processing
is used, the square of the signal to noise ratio for a broadband
measurement of the SB is\cite{explainSNR}
\begin{equation}
\label{snr}
{S^2 \over N^2} = \left( {4 G \rho_c \over 5 \pi c^2} \right)^2 2
{\hat \tau} \int_0^\infty df \, {\Omega_g(f)^2 \gamma(f)^2 \over f^6
S_n(f)^2}.
\end{equation}
Here $\hat \tau$ is the duration of the measurement, and $S_n(f)$ is
the spectral noise density in either detector.  The key quantity
appearing in this equation is the dimensionless function $\gamma(f)$,
which we call the {\it overlap reduction function}.  It characterizes
the reduction in sensitivity to the SB of the detector pair at
frequency $f$ that is due to their separation and non-optimal
orientations, and its value is unity for coincident, aligned
detectors.  In Refs.~\cite{mi,paper} a formula for the overlap
reduction function was derived, which expresses it as an integral over
all solid angles of the complex phase lag between the detectors,
weighted by combinations of the detector beam pattern functions
[cf.~Eq.~$(\ref{gamma_def})$ below].  Christensen\cite{thesis,paper}
numerically calculated this function for various detector
configurations and discussed some of its properties.  However its
dependence on the detector orientations was not apparent.

In this paper we derive an analytic formula for the overlap reduction
function.  Using this formula we are able to carry through steps (iii)
and (iv) outlined above.  We also determine how good are the choices
that have been made for the orientations of the detectors in LIGO,
VIRGO and GEO (an as-yet-unfunded British/German collaboration); i.e.,
we determine how their sensitivity to the SB compares to the
sensitivity they would have if they were optimally oriented.

\subsection{Effect of detector orientations}

Our results are as follows.  Call $\sigma_1$ the angle between the
bisector of the arms of the first detector and the arc of the great
circle that joins the detectors, and similarly define $\sigma_2$ for
the second detector.  Let $\delta = (\sigma_1 - \sigma_2)/2$ and
$\Delta = (\sigma_1+ \sigma_2)/2$, so that $\delta$ describes the
relative rotation of the detector pair, and $\Delta$ describes their
average rotation with respect to the line joining them.  Then in
Sec.~\ref{optimization} below we show that the optimum SNR
$(\ref{snr})$ is given by
\begin{eqnarray}
{S^2 \over N^2} = A \cos^2(4 \delta) + 2 B \cos (4
\delta) \cos( 4 \Delta)  \nonumber \\
\mbox{} + C \cos^2(4 \Delta),
\end{eqnarray}
where the quantities $A$, $B$, and $C$ are independent of $\delta$ and
$\Delta$, and $A$ and $C$ are positive.  There are thus two
possibilities for the optimum detector orientation, depending on the
sign of $B$: (I) $\cos(4 \delta) = - \cos(4 \Delta) = \pm 1$,
corresponding to each detector having an arm along the line joining
them, and (II) $\cos(4 \delta) = \cos(4 \Delta) = \pm 1$,
corresponding to each detector arm being at an angle of $45^\circ$
(mod $90^\circ$) to this line.  In Fig.~\ref{beta} we plot the SNR for
both of these choices of orientation, as a function of the angle
$\beta$ subtended between the detectors at the center of the earth.
The configuration II is optimal for $\beta \alt 70^\circ$, while
configuration I is optimal for detectors which are further apart.
Fig.~\ref{beta} shows that detectors which are close together are the
most sensitive; the sensitivity of LIGO is roughly $\sim 25\%$ of what
it would be if its detectors were coincident.  It also shows that
detectors whose planes are roughly perpendicular ($\beta \sim
90^\circ$) have poor sensitivity, as we would expect.

We also show in Sec.~\ref{optimization} that the narrowband
sensitivity of the detector pair near a given frequency $f$, which is
proportional to $|\gamma(f)|$, is also always optimized at either
configuration I or II.  For example, at very low frequencies,
$|\gamma(f)|$ becomes essentially the overlap of the polarization
tensors of the two detectors [cf.~Eq.~(\ref{general_result}) below
with $\rho_2(0) = \rho_3(0) = 0$], which is maximized in configuration
I.  This low frequency limit was previously derived by Christensen
\cite{paper}.  Motivated by this, he suggested that configuration I
was always the best orientation to choose.  Fig.~\ref{beta} shows
that, though this is not true for some values of the separation angle
$\beta$, the amount lost by choosing I rather than II is never more
than a few percent in SNR.

We now discuss the orientations that have been chosen for the
detector systems that are under construction or that have been
proposed.  The relative rotation angle $\delta$ for a pair of
detectors essentially determines whether the detectors respond to
different polarization components, or to the same polarization
component, of the gravitational wave field.  The advantage in
responding to different components ($2 \delta \sim 45^\circ$) is that
more information can be extracted from incoming burst signals.  On the
other hand, if the detectors respond to the same component ($\delta
\sim 0^\circ$) then the detection signal to noise threshold is reduced,
i.e.~the fact that the same waveform is seen in both detectors means
that one can be more confident that a candidate event is not due to
detector noise.  These considerations guided the choices of the
presently planned values of $\delta$ for the detector pairs LIGO/LIGO
(there will be two LIGO detectors) and VIRGO/GEO, which are $\delta
\approx 0^\circ$ and $2 \delta \approx 45^\circ$ respectively \cite{Kip}.

Within the context of these constraints, a key issue that we wanted to
understand was the following: given the above values of $\delta$, how
much does the broadband sensitivity of the detector pair depend on
$\Delta$, i.e.~by how much can the SNR be reduced if $\Delta$ is
chosen arbitrarily instead of being optimally chosen.  The answer we
obtain [Secs.~\ref{gamma_sec} and \ref{optimization} below] is that
for relatively close detectors with $\beta \alt 30^\circ$, the
dependence on $\Delta$ is very weak; but the dependence is strong for
detectors on different continents.  Hence for LIGO's parallel
detectors, the sensitivity will be close to optimal irrespective of
the value of $\Delta$; the present choice of $\Delta = 28.2^\circ$
implies that the SNR for LIGO is $\sim 97\%$ of the sensitivity at
optimal orientation.  The VIRGO/GEO orthogonal detector pair, however,
will have a SNR of less than $10^{-3}$ times the optimal value,
irrespective of the value of $\Delta$, if $2 \delta$ is chosen to be
$45^\circ$ (as is planned, so as to optimize the information
obtainable from burst sources).

Finally, we estimate the $90\%$ upper confidence limit that can be
placed on $\Omega_g(f)$ by the so called ``advanced detectors'' in
LIGO \cite{ligo_science}, in a broadband measurement using a one third
of a year integration time, to be $\sim 5 \times 10^{-10}$ in the
frequency band $20 \, {\rm Hz} \alt f \alt 70 \, {\rm Hz}$
[cf.~Eq.~$(\ref{expected})$ below].  This is a little worse than
earlier estimates which assume that the detectors are coincident and
aligned \cite{300years,prospect}.

\subsection{Organization of this paper}

The layout of this paper is as follows.  In Sec.~\ref{correlations} we
define the cross correlation matrix for a network of detectors and
give the formula for the contribution to this quantity from the SB.
In Sec.~\ref{processing} we describe the general optimal
data-processing strategy, the justification of which is given in
Appendix \ref{stats}.  In this appendix we also derive the
generalization of the signal to noise formula for optimal signal
processing $(\ref{snr})$, to a network of more than two detectors, but
more importantly to a network that has more than one interferometer
per site: we show how to take into account the effect of correlated
noise in detectors at the same site by introducing the concept of the
{\it effective spectral noise density} of a detector site.  In
Appendix \ref{gamma_calc} we derive the formula for the overlap
reduction function, and we describe some of its properties in
Sec.~\ref{gamma_sec}.

Next, in Sec.~\ref{optimization}, we show how to optimize the
orientations of a pair of detectors, both for narrowband and for
broadband measurements of the SB.  Section \ref{implications}
describes the implications of our results for the LIGO, VIRGO and GEO
detector facilities.  Finally in Sec.~\ref{conclusion} we summarize
our main results.

We use units throughout in which the speed of light $c$ and Newton's
gravitational constant $G$ are unity.

\section{DETECTOR CROSS CORRELATION MATRIX}
\label{correlations}

The effect of a stochastic background on a detector network will be
essentially to produce statistical correlations between the outputs of
the various detectors.  A key result which we will need is an
expression for these correlations in terms of the spectrum
$\Omega_g(f)$ of the gravitational background.  This was first given
by Christensen\cite{thesis,paper}, although it is implicit in the work
of Michelson\cite{mi}.  We briefly describe the derivation in this
section, and we lay the foundations for our analysis of Appendix
\ref{stats} and Sec.~\ref{processing}.

A detector network with $N$ detectors will have outputs
\begin{equation}
\label{decomposition_h}
h_a(t) = h_a^{\rm signal}(t) + n_a(t) + s_a(t),
\end{equation}
for $1 \le a \le N$.  Here $h_a$ is the strain amplitude that we read
out from the $a {\rm th}$ detector; it consists of an intrinsic
detector noise $n_a$, a contribution from the SB $s_a$, and possibly a
contribution $h_a^{\rm signal}$ from non-stochastic gravitational
waves (bursts and periodic waves).  The noise $n_a$ and the SB induced
strain $s_a$ are independent random processes, which we assume to be
gaussian and stationary.

The detector correlations can be described by the correlation matrix
\FL
\begin{equation}
\label{c_h_def}
C_h(\tau)_{ab} = \langle h_a(t + \tau) h_b(t) \rangle -
\langle h_a(t + \tau) \rangle \, \langle h_b(t) \rangle,
\end{equation}
where $\langle \dots \rangle$ means an ensemble average or a time
average.  The Fourier transform of the correlation matrix, multiplied
by two, is the power spectral density matrix:
\begin{equation}
\label{s_h_def}
S_h(f)_{ab} = 2 \int_{-\infty}^{\infty} d \tau \, e^{2 \pi i f \tau}
C_h(\tau)_{ab}.
\end{equation}
This is a positive definite hermitean matrix which satisfies the equations
\begin{equation}
\label{p1}
\langle {\tilde h}_a(f) {\tilde h}_b(f^\prime)^* \rangle = {1 \over 2}
\delta(f - f^\prime) S_h(f)_{ab},
\end{equation}
and
\FL
\begin{equation}
\label{p2}
\left< e^{  i \int dt \, w_a(t) h_a(t) } \right> =
\exp \left\{ - {1 \over 2} \int_0^\infty d f {\tilde {\bf w}}^\dagger
\cdot {\bf S}_h \cdot {\tilde {\bf w}} \right\},
\end{equation}
for any functions $w_a(t)$.
Here tildes denote Fourier transforms, according to the convention that
$$
{\tilde h}(f) = \int e^{2 \pi i f t} h(t) dt.
$$
Since the random processes $n_a(t)$ and $s_a(t)$ are uncorrelated, the
spectral density matrix of the detector outputs is just the sum of
those for the detector noise and for the background:
\begin{equation}
\label{sn_sum}
{\bf S}_h(f) = {\bf S}_n(f) + {\bf S}_s(f).
\end{equation}

To derive an expression for ${\bf S}_s(f)$, two key ingredients are
needed.  The first is a mode expansion for the metric perturbation for
an isotropic, stationary SB.  Expressed in frequency space, this is
\FL
\begin{equation}
\label{mode_sum_2}
{}^{\rm (SB)}{\tilde h}_{ij}({\bf x},f) = \int d^2 \Omega_n
\,\sum_{A = +,\times} \,s_{A,{\bf n}}(f) \, e^{2 \pi i f {\bf n} \cdot
{\bf x}} \, e_{ij}^{A,{\bf n}}
\end{equation}
for $f \ge 0$, and ${}^{\rm (SB)} {\tilde h}_{ij}({\bf x},f) = {}^{\rm
(SB)}{\tilde h}_{ij}({\bf x},- f)^*$ for $f < 0$.  Here the tensors
${\bf e}^{A,{\bf n}}$ are the usual transverse traceless polarization
tensors, normalized according to $e_{ij}^{A,{\bf n}} e_{ij}^{B,{\bf
n}} = 2 \delta_{AB}$, and $\int d^2 \Omega_n$ denotes the integral
over solid angles parameterized by the unit vector ${\bf n}$.  The
coefficients $s_{A,{\bf n}}$ are random processes which satisfy
\cite{correction}
\FL
\begin{eqnarray}
\label{s_moments}
\langle \,s_{A,{\bf n}}(f) \,s_{B,{\bf m}}(f^\prime)^* \, \rangle
&=& \delta_{AB} \,\delta^2({\bf n}, {\bf m})  \nonumber \\
\mbox{} & & \times \, \delta(f - f^\prime) \,\,{\rho_c \over 4 \pi
f^3}\,\, \Omega_g(f)
\end{eqnarray}
and
\begin{equation}
\label{s_moments1}
\langle \,s_{A,{\bf n}}(f) \,s_{B,{\bf m}}(f^\prime) \, \rangle = 0,
\end{equation}
for $f,f^\prime \ge 0$ \cite{relic}.  Here $\delta^2({\bf n},{\bf m})$
is the delta function on the unit sphere.

The second ingredient is the expression for the response of the $a$th
detector to the background.  This is
\begin{equation}
\label{response}
s_a(t) = {\bf d}_a : {}^{\rm (SB)}{\bf h}({\bf x}_a,t),
\end{equation}
where ${\bf x}_a$ is the position of the detector, the $:$ denotes a
double contraction, and ${\bf d}_a$ is a symmetric tensor that
characterizes the detector's orientation (its polarization tensor).  If
the arms of the detector are in the directions of the unit vectors
${\bf l}$ and ${\bf m}$, then ${\bf d}_a = ({\bf l} \otimes {\bf l} -
{\bf m} \otimes {\bf m})/2$ \cite{Forward}.  From
Eqs.~$(\ref{mode_sum_2})$ and $(\ref{response})$ we obtain that
\begin{equation}
{\tilde s}_a(f) = \sum_A \int d^2 \Omega_n \, F_a^A({\bf n}) \, s_{A,{\bf
n}} \, e^{2 \pi i f {\bf n} \cdot {\bf x}_a},
\end{equation}
where $F_a^A({\bf n}) \equiv {\bf d}_a : {\bf e}^{A,{\bf n}}$ are the
detector beam pattern functions.  Inserting this
response function into an equation analogous to Eq.~$(\ref{p1})$, and
using Eqs.~$(\ref{s_moments})$ and $(\ref{s_moments1})$, we obtain
\begin{equation}
\label{correlation}
S_s(f)_{ab} = {4 \rho_c \Omega_g(f) \over 5 \pi f^3 } \, \gamma_{ab}(f),
\end{equation}
where \cite{equiv_results}
\begin{eqnarray}
\label{gamma_def}
\gamma_{ab}(f) = {5 \over 8 \pi} \int d^2 \Omega_n \, (F_a^+ F_b^+ +
F_a^\times F_b^\times) \nonumber \\
 \mbox{} \times \exp \left[ 2 \pi i f {\bf n} \cdot ({\bf x}_a - {\bf
x}_b) \right].
\end{eqnarray}

The functions $\gamma_{ab}$ are the overlap reduction functions
discussed in Sec.~\ref{intro}.  It can be seen that their value is
unity for coincident, aligned detectors.  Below when considering a
single pair of detectors we shall write $\gamma_{ab}(f)$ simply as
$\gamma(f)$.

\section{THE OPTIMAL PROCESSING STRATEGY}
\label{processing}

In this section we describe the optimal method for filtering the
detector outputs when searching for the SB.  We discuss the two
detector case in subsection \ref{two}; the filtering method in this
subsection is implicit in the formulae of Ref.~\cite{mi}.  A detailed
proof that the method is optimal is given in Appendix \ref{stats}.  In
subsection \ref{cnoise} we discuss the effects of correlated sources
of noise, and explain why correlation measurements between detectors
at widely separated sites yield much better upper bounds on
$\Omega_g(f)$ than correlation measurements between detectors at one
site.  Next we describe the modifications to the filtering method
necessitated by correlated noise, in subsection \ref{extra}.

\subsection{General description}
\label{two}

To measure the stochastic background one really needs to measure the
spectral density matrix of the detector outputs ${\bf S}_h(f)$.  Now
there is no way in principle to separate out the portions of ${\bf
S}_h(f)$ due to detector noise ${\bf S}_n(f)$, and due to the SB ${\bf
S}_s(f)$.  If we had only one detector, we could only conclude that
$S_s(f) \le S_h(f)$.  From Eq.  $(\ref{correlation})$ and using
$\gamma_{aa}(f) = 1$, this would give an upper bound on $\Omega_g$ of
\begin{equation}
\label{est}
\Omega_g(f) \alt 2.5 \times 10^{-6} \left( {h_n(f) \over 10^{-23} }
\right) \left( {f \over 100 \, {\rm Hz}} \right) h_{75}^{-2}.
\end{equation}
Here $h_n = \sqrt{f S_h(f)}$, which is projected to be $\agt 10^{-23}$
for LIGO, even at an advanced stage \cite{ligo_science}.  The quantity
$h_{75}$ is the Hubble constant scaled to the value of $75 \, {\rm km
\, s^{-1} Mpc^{-1}}$.

It is unlikely that $\Omega_g$ will be as large as the value in
Eq.~$(\ref{est})$.  However, if it does happen that $\Omega_g \agt
10^{-5}$ in the LIGO/VIRGO waveband, then the SB induced noise may
dominate over the other sources of detector noise at some frequencies,
and may ultimately constrain the amount of information that we can
extract from burst gravitational waves.  In this paper we
shall from now on assume that $\Omega_g$ is small, and restrict
attention to measurements made using two or more detectors.

With two or more detectors, one takes advantage of the fact that the
sources of noise in each detector will be independent.  This will be
the case for detectors at widely separated sites, because sources of
noise that are correlated between the detectors on timescales of the
order of the light travel time between them are expected to be
insignificant, or if not they can be monitored and compensated
for \cite{paper}.  When correlated noise is unimportant, then the
off-diagonal elements of ${\bf S}_n(f)$ will be very small, so that
\begin{equation}
S_s(f)_{ab} \approx S_h(f)_{ab} \,\,\, \mbox{ for } a \ne b.
\end{equation}
By measuring these components we can gain information about the SB.
One does this by cross correlating the two output streams
\cite{300years}.  One takes each detector output $h_a(t)$, $a = 1,2$,
and constructs using an optimizing linear filter $K(t)$ the quantity
${\tilde H}_a(f) = {\tilde K}(f) {\tilde h}_a(f)$.  The purpose of
this filter is essentially to suppress the signal at those frequencies
at which the detector noise is strong, and it is given by [cf.
Eq.~(\ref{Hfilter}) below]
\begin{equation}
{\tilde K}(f) = {1 \over f^{3/2} S_h(f)}.
\end{equation}

For coincident, aligned detectors the next step is simply to integrate
$H_1(t)$ against $H_2(t)$, see, e.g., Ref.~\cite{300years}.
For noncoincident detectors, however, a different strategy is
necessary.  One first constructs the correlation with time delay $\tau$,
\begin{equation}
Y(\tau) = \int_{-{\hat \tau} /2}^{{\hat \tau}/2} dt \, H_1(t+\tau)
H_2(t),
\end{equation}
where ${\hat \tau}$ is the observation time, typically of the order of
a year.  Then one calculates the weighted average
\begin{equation}
Y = \int_{-\tau_1}^{\tau_1} d\tau \, L(\tau) Y(\tau),
\end{equation}
where $\tau_1$ is the light travel time between the detectors, and
$L(\tau)$ is a weighting function which must be carefully chosen for
each detector pair in order to maximize the sensitivity.  Roughly
speaking, this smearing of the cross-correlation compensates in some
measure for the phase lags between the detectors which were discussed in
Sec.~\ref{intro}.  The quantity $Y$ will then have a truncated
gaussian distribution (i.e. gaussian but restricted to positive
values) with signal to noise ratio given by Eq.~$(\ref{snr})$
\cite{explainSNR}.

In fact the sliding delay function $L(\tau)$ is just the Fourier
transform of the overlap reduction function [see
Eqs.~(\ref{omega_est}) and (\ref{Hfilter})].  In Appendix
\ref{gamma_calc} we give an analytic formula for $L(\tau)$, and we
show in Fig.~\ref{delay} the sliding delay function that will need to
be used for the LIGO pair of detectors.

In order of magnitude, the $90\%$ confidence upper limit that can be
placed on $\Omega_g$ by cross-correlating between the detectors is
\cite{300years}
\begin{equation}
\label{win_factor}
\Omega_g^{\rm max} \approx {\Omega_0 \, \over \sqrt{{\hat \tau} \Delta f}}.
\end{equation}
Here $\Omega_0 \sim 10^{-6}$ is the upper bound (\ref{est}) obtainable
from one detector, and $\Delta f$ is the bandwidth of the measurement.
[If no bandpass filtering of the data is carried out, $\Delta f$ will be
roughly the width of the peak of the function $1 / (f^3 S_n )$; and if
filtering with a bandwidth $\Delta f$ is used, then the domain of
integration in Eq.~(\ref{snr}) must be suitably restricted].

\subsection{Effects of correlated noise.}
\label{cnoise}

Up to this point we have assumed that the intrinsic detector noise is
uncorrelated between different detectors, i.e.~that the matrix ${\bf
S}_n(f)$ is diagonal.  We now relax this assumption and consider the
effects of correlated noise.  It is planned for LIGO/VIRGO to
ultimately have two or three detectors per detector site, perhaps
optimized for different types of gravitational wave sources.  Thus,
there will be two possible types of correlation measurements: {\it
intrasite} measurements at one site, and {\it intersite} measurements
between detectors at different sites.  As previously mentioned, it
seems very unlikely that there will be any significant correlated
noise in intersite measurements, and so it will only be important for
intrasite measurements.

We now consider what information can be extracted from intrasite
correlations.  The quantity we can measure is the sum ${\bf S}_n(f)+
{\bf S}_s(f)$, the two terms of which are in principle
indistinguishable \cite{disclaimer}.  Hence, we can draw inferences about
${\bf S}_s(f)$ {\it only} if we have some information about the
correlated noise.  The most obvious such a priori information is the
fact that ${\bf S}_n$ is positive hermitean matrix.  It follows that
if we measure the total power spectral density matrix to be ${\hat {\bf
S}}_h$, then ${\bf S}_s < {\hat {\bf S}}_h$.  However, this only tells
us that $\Omega_g \agt 10^{-6}$, [cf.~Eq.~(\ref{est}) above], since
the inequality applies to the total matrix, and the off-diagonal
elements are small compared to the diagonal elements.  In particular,
it is not true without further assumptions that
\begin{equation}
S_s(f)_{ab} \, \le \, |\, {\hat S}_h(f)_{ab}\,| \,\,\,\,\,\,\,\,\,\,
\mbox{ for } a \ne b,
\end{equation}
(as is implicitly assumed in Ref.~\cite{paper}), since $S_n(f)_{ab}$
may be negative (or complex).  A complex value of $S_n(f)_{ab}$
corresponds physically to sources of noise that excite two detectors
with a certain preferred phase lag between them.

In Appendix \ref{stats}, we derive the $90\%$ confidence upper limit
that one can place on $\Omega_g$ in a bandwidth $\Delta f$ using
intrasite correlations, which generalizes Eq.~(\ref{win_factor}).  If
we define the noise correlation coefficients by
\begin{equation}
{\cal C}_{ab} \equiv { S_n(f)_{ab} \over \sqrt{ S_n(f)_{aa}
\, S_n(f)_{bb} } },
\end{equation}
then the result depends on (i) the measured value ${\hat {\cal C}}$ of
${\cal C}$ (determined from ${\hat {\bf S}}_h$), and (ii) the assumed
a priori maximum value ${\cal C}_{\rm max}$ of $| {\cal C}|$.  In
order of magnitude we find [cf.~Eqs.~(\ref{beta2}) and (\ref{pc2}) below]
\begin{equation}
\label{win_factor1}
\Omega_g^{\rm max} \approx \Omega_0 \, \left[ {\hat {\cal C}} + \sqrt{
{1 \over {\hat \tau} \Delta f } + {\cal C}_{\rm max}^2 } \, \right].
\end{equation}
Since there will be various unknown sources of weak correlated noise,
it is not appropriate to choose a very small value of ${\cal C}_{\rm
max}$.  Hence, the upper bound (\ref{win_factor1}) will be much worse
than the bound (\ref{win_factor}) obtained from intersite
correlations.

We also consider the assumption that the correlated detector noise
always excite different detectors in phase, but can
be arbitrarily large in magnitude, so that
\begin{equation}
\label{your_kidding}
S_n(f)_{ab} \ge 0  \mbox{    for } a \ne b.
\end{equation}
In this case intrasite correlation measurements cannot be used to {\it
detect} the SB, but can only be used to place upper bounds on its
magnitude.  The resulting upper limit [cf.~Eq.~(\ref{pc1})] is given
by Eq.~(\ref{win_factor1}) with ${\cal C}_{\rm max} = 0$.  Thus, the
intersite correlations will still give better bounds unless the actual
amount of correlated noise in the detectors satisfies
\begin{equation}
\label{condt}
{\hat {\cal C}} \alt {1 \over \sqrt{{\hat \tau} \Delta f} }
\end{equation}
which is $\sim 10^{-4}$ for a year-long measurement with a bandwidth
of $\sim 50 \, {\rm Hz}$.  Since it is not clear that either of the
conditions (\ref{your_kidding}) or (\ref{condt}) will be appropriate,
we henceforth consider only intersite correlations.

\subsection{Filtering intersite correlation measurements.}
\label{extra}

In Appendix \ref{stats} we show how to optimally filter the intersite
data when there is a detector network with several interferometers per
detector site.  In this case the detector noise matrix ${\bf S}_n(f)$
will have a block diagonal form, with each block corresponding to a
site.  Let ${\bf S}_A$ be the subblock corresponding to the $A$th
site.  The strategy is essentially to measure the off-diagonal blocks
of ${\bf S}_h(f)$ (i.e.~the intersite correlations), and to use these
to obtain information about the SB.  The resulting SNR squared is then
given [cf.~Eq.~(\ref{snr3}) below] by a simple generalization of
Eq.~$(\ref{snr})$:
\begin{eqnarray}
\label{snr1}
{S^2 \over N^2} = \left( {4 \rho_c \over 5 \pi} \right)^2 2
{\hat \tau} & & \int_0^\infty df \, {\Omega_g(f)^2 \over f^6}
\nonumber \\
\mbox{} & & \times \sum_{A < B} {\gamma_{AB}(f)^2 \over S_A^{\rm
(eff)}(f) S_B^{\rm (eff)}(f)}.
\end{eqnarray}
Here the sum is over the pairs of sites, and $\gamma_{AB}$ is the
overlap reduction function for any detector at site $A$ together with
any detector at site $B$.  The quantities $S_A^{\rm (eff)}(f)$, which
we call the effective site spectral noise densities for correlations
between sites, are given by
\begin{equation}
\label{site_noise0}
S_A^{\rm (eff)}(f) = \left[ \sum_{ab} \left( {\bf S}_A^{-1}
\right)_{ab} \right]^{-1}.
\end{equation}
They will be real since the matrices ${\bf S}_A$ are Hermitean.  In the
case where the off-diagonal elements of ${\bf S}_A$ are all equal to
$S_n^\prime(f)$, and the common value of the diagonal elements is
$S_n(f)$, then we obtain that
\begin{equation}
\label{site_noise}
S_A^{\rm (eff)}(f) = {1 \over N} S_n(f)  + (1 - {1 \over N})
S_n^\prime(f),
\end{equation}
where $N$ is the number of detectors.  This formula shows that the use
of $N$ detectors instead of a single detector at one site reduces that
sites effective noise density by a factor of $1/N$, {\it if} the
detectors noise sources are independent ($S_n^\prime \approx 0$).  If
the noise sources are strongly correlated, however, so that
$S_n^\prime \approx S_n$, then there is no significant reduction in
the effective noise.

\section{The overlap reduction function}
\label{gamma_sec}

In the preceding sections we have seen that for a pair of
interferometric detectors, the overlap reduction function $\gamma(f)$
characterizes the dependence on the detector separation and
orientations of both the correlation matrix $(\ref{correlation})$, and
the broadband sensitivity to the SB $(\ref{snr})$.  We now give an
analytic formula for this function and discuss some of its properties.

\subsection{The general formula for $\gamma(f)$}

We first define the variables which describe the orientations and
separation of a pair of detectors.  Let $L$ be the line joining the
two detectors and $P_1$ ($P_2$) be the plane formed by the arms of the
first (second) detector; see Fig.~\ref{angles}.  The variables we will
use are: (i) The distance $d$ between the detectors; (ii) The acute
angle $\beta_1$ between $L$ and $P_1$; (iii) The angle $\sigma_1$
between the projection of $L$ onto $P_1$ and the bisector of the two
arms of the first detector; (iv) Corresponding angles $\beta_2$ and
$\sigma_2$; (v) The angle $\chi$ between $L$ and and the intersection
of $P_1$ and $P_2$.  The directions (clockwise or anticlockwise) in
which $\sigma_1$ and $\sigma_2$ are chosen to be positive are
unimportant, as long as the conventions for $\sigma_1$ and for
$\sigma_2$ coincide as $\beta_1, \beta_2 \to 0$.  Let $\delta \equiv
(\sigma_1 - \sigma_2)/2$ and $\Delta =( \sigma_1 + \sigma_2)/2 $.  In
Sec.~\ref{intro} we have called these the {\it relative rotation} and
{\it total rotation} angles respectively, since for $\beta_1 = \beta_2
=0$, the angle $\delta$ is half of the relative rotation of the
detectors while $\Delta$ measures the average rotation with respect to
the line joining them \cite{commenta}.

In general the overlap reduction function as given in
Eq.~$(\ref{gamma_def})$ will depend on all of the variables $\beta_1$,
$\beta_2$, $\delta$, $\Delta$, $\chi$, and on the phase lag $\alpha =
2 \pi f d $ between the detectors, where $f$ is the frequency.  Now for
terrestrial detectors $\chi = \pi/2$ and $\beta_1 = \beta_2$ ($=
\beta$ say), which is also the angle subtended at the center of the
earth by the detector pair.  For this case we derive in Appendix
\ref{gamma_calc} the following formula:
\begin{equation}
\label{gamma_general}
\gamma(f) = \cos (4 \delta) \Theta_1(\alpha,\beta) + \cos(4 \Delta)
\Theta_2(\alpha,\beta),
\end{equation}
where the functions $\Theta_1$ and $\Theta_2$ are
\begin{equation}
\Theta_1(\alpha,\beta) = \cos^4 (\beta /2) g_1(\alpha)
\end{equation}
and
\begin{eqnarray}
\label{theta2}
\Theta_2(\alpha,\beta) &=& \cos^4 (\beta/2) g_2(\alpha) + g_3(\alpha)
\nonumber \\
\mbox{} & &  - \sin^4 (\beta/2) \left[ g_2(\alpha) + g_1(\alpha) \right].
\end{eqnarray}
The functions $g_j(\alpha)$ are given in Eqs.~(\ref{g1}),
(\ref{g2}) and (\ref{g3}) of the Appendix, and are all linear
combinations of the functions $\sin (\alpha) / \alpha^n$ and $\cos
(\alpha) / \alpha^{n}$ where $1 \le n \le 5$.

Several properties of the overlap reduction function $\gamma(f)$,
which were discovered by Christensen and others from numerical
calculations\cite{thesis,mi,paper}, can be read off the formula
$(\ref{gamma_general})$.  First, there will always be frequencies $f$
for which $\gamma(f)$ vanishes, and correspondingly near which the
narrow-band sensitivity of the detector pair to the SB is very poor.
For detectors that are less than a few thousand kilometers apart, the
first null frequency is at $f_1 \sim (70 \, {\rm Hz}) (3000 \, {\rm
km} / d)$, irrespective of the detector orientations (see below).
Second, the reduction function falls off like $1/f$ when $\alpha \gg
1$, or equivalently when $f \gg f_1$.  Hence the 90\% confidence limit
that we can place on $\Omega_g(f)$ scales like $1/(f d)$ for large
$d$.  However for detectors which are $\sim 3000 \, {\rm km}$ apart,
the phase lag $\alpha$ is of order unity for typical detector
frequencies, and hence the variation of the sensitivity with distance
is more complex than simply scaling like $1/d$ (see
Sec.~\ref{implications} below and also Fig.~\ref{beta} above).

\subsection{Some special cases}
\label{special_cases}

To understand the behavior of $\gamma(f)$ as a function of $\delta$,
$\Delta$ and $\beta$, it is useful to consider some limiting cases.
For coplanar, coincident detectors, $d = \beta = 0$, and using $g_1(0)
= 1$ and $g_2(0) = g_3(0) = 0$ we find that $\gamma = \cos(4
\delta)$.  This is just what we would expect physically.  Consider,
for instance, detectors which are rotated with respect to one another
by $45^\circ$.  Vertically incident gravitational waves will couple to
the two detectors via polarization components that are orthogonal and
hence statistically independent; thus these waves give no contribution
to the cross-correlation.  Waves which are incident from non-vertical
directions do give rise to correlations between the detectors, but
when we average over all incident directions the total correlation
vanishes \cite{average_directions}.  It is also clear from rotational
symmetry that there should be no dependence on the total rotation
angle $\Delta$.

Now suppose that the detectors are still coincident but no longer
coplanar, so that $\beta \ne 0$.  Equivalently, suppose the detectors
are separated by a distance $d$ but evaluate $\gamma(f)$ at
frequencies $f \ll 1/d$.  The result is
\FL
\begin{equation}
\label{gamma0}
\gamma(f) = \cos^4 (\beta/2) \cos(4 \delta) - \sin^4 (\beta/2) \cos( 4
\Delta).
\end{equation}
This equation agrees with Eq.~(3.9) of Ref.~\cite{paper} after making an
appropriate change of variables.  As we would expect, the dependence on
$\Delta$ is small when $\beta$ is small; it is smaller than the
dependence on the relative rotation angle $\delta$ by a factor of
$\tan^4 (\beta/2)$ ($\sim 3 \times 10^{-3}$ for the planned LIGO
detectors; see below).

A third limiting case is when $\beta = 0 $ but $d \agt 1/f$, so that the
detectors are coplanar but effectively separated.  In this case,
\FL
\begin{equation}
\label{coplanar}
\gamma(f) = \cos (4 \delta) g_1(\alpha) + \cos(4 \Delta) \left\{
g_2(\alpha) + g_3(\alpha) \right\}.
\end{equation}
Plots of the functions $g_1(\alpha)$ and $g_2(\alpha) + g_3(\alpha)$
are shown in Fig.~\ref{planar}.  We see that for $\alpha \alt 4$, the
$\cos(4 \delta)$ term dominates, just as above for coincident
detectors for small $\beta$.  When the phase lag $\alpha$ is large,
the functions in Eq.~$(\ref{coplanar})$ can be approximated as
$g_0(\alpha) \approx 5 \sin \alpha / (16 \alpha)$ and $g_2(\alpha) +
g_3(\alpha) \approx - g_0(\alpha)$, so we obtain
\begin{equation}
\gamma(f) \approx - {5 \sin \alpha \over 8 \alpha} \sin( 2 \sigma_1)
\sin (2 \sigma_2).
\end{equation}
What is happening physically here is that the cross-correlation is
dominated by modes whose propagation vectors are nearly parallel to
the line joining the detectors.  This can be seen by applying a
stationary phase argument to the integral $(\ref{gamma_def})$.  Hence
it is reasonable that the overlap reduction function should have
separate factors for each detector that depend on how they are
oriented with respect to the line joining the detectors.  In this case
the dependence of the overlap reduction function $\gamma$ on the total
rotation angle $\Delta$ is strong.

In summary, the dependence on the relative rotation angle $\delta$
will dominate over the dependence on the total rotation angle $\Delta$
unless two conditions are satisfied: (i) $\beta \agt \pi/4$, and (ii)
$\alpha \agt 1$, i.e.~$d \agt c / f_n$, where $f_n \sim 100 \, {\rm
Hz}$ is a typical frequency at which the detector is sensitive.  In
reality of course $\beta$ and $d$ are not independent; they are
related by
\begin{equation}
\label{distance}
d = 2 \sin (\beta/2) r_E,
\end{equation}
where $r_E$ is the radius of the earth.  Due to the coincidence that
the quantity $f_n r_E / c$ is of order unity, conditions (i) and (ii)
above become satisfied at approximately the same value of $\beta$.

For a pair of terrestrial detectors, as $\beta$ and $d$ are increased
from zero, the following three effects occur: (i) $\gamma(f)$
decreases because $d$ is increasing, as discussed above (ii)
$\gamma(f)$ decreases because $\beta$ is increasing.  From
Eq.~$(\ref{gamma0})$ we see that the maximum over all orientations of
the value of $\gamma(0)$ is $1 - \sin^2 \beta /2$.  Hence the effect
of increasing $\beta$ contributes typically a factor of at most $1/2$
to the decrease in $\gamma(f)$.  (iii) The
dependence of $\gamma(f)$ on $\Delta$ increases to become as strong as
the dependence on $\delta$.

\section{Optimization of detector orientations}
\label{optimization}

We now turn to the issue of how to {\it optimize} the orientations of
a single pair of detectors so as to detect or measure the SB.  We
assume that the detector's noise curves are the same, so that
$S_n(f)_{aa} = S_n(f)_{bb} = S_n(f)$.  We consider separately the
cases of maximizing the narrowband sensitivity to the SB in the
vicinity of some frequency $f$, and of maximizing the overall
broadband sensitivity.

Now if we rotate a detector through $90^\circ$, then its polarization
tensor ${\bf d}$ [cf.~Eq.~$(\ref{response})$] will have its sign
flipped.  Hence $|\gamma(f)|$ is a periodic function of both of the
angles $\sigma_1$, $\sigma_1$ (or $\delta$, $\Delta$ ) with period
$90^\circ$.  In the following discussion, all values of these angles
and all equations involving them should be treated modulo $90^\circ$.

\subsection{Narrowband sensitivity}

 From Eq.~$(\ref{snr})$ we see that the $90\%$ confidence limit that we
can place on $\Omega_g(f)$ in a small interval of frequency of width
$\Delta f$ centered about $f$, using a measurement of duration $\hat
\tau$, is \cite{explainSNR}
\begin{equation}
\label{om90_narrow}
\Omega_g^{(90\%)}(f;\Delta f) = \left( { 5 \pi \over 4 \rho_c }
\right) {f^3 S_n(f) \over |\gamma(f)| } {1.65 \over \sqrt{2 {\hat
\tau} \Delta f}}.
\end{equation}
To minimize this we need to maximize the quantity $|\gamma(f)|$.  From
Eq.~$(\ref{gamma_general})$ it follows that the maximum value over all
orientations of $|\gamma(f)|$ is $|\Theta_1(f)| +
|\Theta_2(f)|$.

Now if the signs of $\Theta_1$ and $\Theta_2$ are different, then this
maximum will be achieved at $\sigma_1 = \sigma_2 = 45^\circ$
(i.e.~either $\delta = 45^\circ$, $\Delta = 0^\circ$ or $\delta = 0^\circ$,
$\Delta = 45^\circ$).  This corresponds to each detector having one
arm parallel to the line which joins them.  We will call this type of
orientation {\it configuration I}.  This will be the optimal
orientation at zero frequency, since $\Theta_1(f=0) = \cos^4 (\beta/2)
> 0$, and $\Theta_2(f=0) = - \sin^4 (\beta/2) < 0$.  It will also be
optimal at high frequencies ($fd \gg 1$), since from from
Eqs.~$(\ref{gamma_general})$ and (\ref{g1}) -- (\ref{fdef}) we see
that $\Theta_1(\alpha) \approx - \Theta_2(\alpha)$ for $\alpha \gg 1$.

If, on the other hand, the signs of $\Theta_1$ and $\Theta_2$ are the
same, then we see from Eq.~$(\ref{gamma_general})$ that $|\gamma(f)|$ will be
maximized at $\sigma_1 = \sigma_2 = 0^\circ$ (i.e.~either $\delta = \Delta =
0^\circ$ or $\delta = \Delta = 45^\circ$).  In this case, which we will call
{\it configuration II}, each detector arm makes an angle of $45^\circ$
(mod $90^\circ$) to the line joining the detectors.  This will be
optimal, e.g., for $10 \, {\rm Hz} \alt f \alt 40 \, {\rm Hz}$ when
$\beta = 29^\circ$ (the LIGO value; see below).

Now suppose that the detector pair is in either configuration I or
configuration II.  Then for some frequencies the orientations will be
optimal, while for some other frequencies they will not be.  It is
useful to calculate, for a given frequency $f$, the factor ${\cal
R}(f)$ by which $|\gamma(f)|$ is reduced if the wrong configuration
out of I and II is chosen.  Note that choosing between I and II is
equivalent to fixing the relative rotation angle $\delta$ to be zero,
and then choosing $\Delta$ to be either $0^\circ$ or $45^\circ$.
Hence ${\cal R}$ measures the sensitivity of $\gamma(f)$ to the total
rotation angle $\Delta$ when the detectors are parallel.  From
Eq.~$(\ref{gamma_general})$, we obtain
\begin{equation}
{\cal R} = { | \, |\Theta_1| - |\Theta_2| \, | \over  |\Theta_1| + |\Theta_2|}.
\end{equation}
At low frequencies ($\alpha = 2 \pi f d \ll 1$) this is approximately
\begin{equation}
{\cal R}(f \to 0) \approx \left( 1 + { 2 \nu^4 \over 1 - 2 \nu^2} \right)^{-1},
\end{equation}
where $\nu = \sin (\beta /2)$.  Thus for small values of $\beta$ we
have ${\cal R}(f \to 0) \approx 1$, and whether configuration I or II
is chosen is relatively unimportant.

At high frequencies ($\alpha \gg 1$), however, we find that ${\cal
R}(f)$ asymptotes to a constant ${\cal R}_\infty$ independent of
frequency, where
\begin{equation}
\label{loose}
{\cal R}_\infty \approx \left( {2 \nu \over \nu^2 + 1} \right)^2,
\end{equation}
which can be quite small.  [${\cal R}_\infty \approx 0.2$ for the LIGO
detectors; see below.]  Hence the choice of $\Delta$ will have a large
effect on the high frequency sensitivity to the SB, especially for
detectors which are close together for which $\nu$ is small.

\subsection{Broadband sensitivity}

It is quite likely that the SB will be so weak that it will only be
detectable (if at all) by integrating over a broad range of
frequencies, e.g.~from $\sim 20 \, {\rm Hz}$ to $\sim 70 \, {\rm
Hz}$.  Hence it is more important to optimize the overall broadband
response $(\ref{snr})$ of the pair of detectors than to optimize the
narrow band sensitivity at some frequency.  We now determine how to do
this if we assume that the spectrum $\Omega_g(f)$ is approximately
constant over the relevant frequency range - the so-called
``Zel'dovich spectrum''.

By inserting Eq.~$(\ref{gamma_general})$ into Eq.~$(\ref{snr})$, we
find that the SNR squared after optimal filtering is
\begin{eqnarray}
\label{snr2}
{S^2 \over N^2} = A(\beta) \cos^2(4 \delta) + 2 B(\beta) \cos (4
\delta) \cos( 4 \Delta)  \nonumber \\
\mbox{} + C(\beta) \cos^2(4 \Delta),
\end{eqnarray}
where
\begin{equation}
\label{diagnostic}
B(\beta) = \left( {4 \rho_c \over 5 \pi } \right)^2 2
{\hat \tau} \int_0^\infty df \, {\Omega_g^2 \Theta_1 \Theta_2
\over f^6 S_n^2},
\end{equation}
and $A$ and $C$ are given by similar formulae with $\Theta_1 \Theta_2$
replaced by $\Theta_1^2$ and $\Theta_2^2$ respectively.  Note that
$A$, $B$ and $C$ depend on $\beta$ in two distinct ways, since from
Eq.~$(\ref{distance})$,
\begin{equation}
\Theta_{1,2}(f) = \Theta_{1,2}( \alpha = 4 \pi f r_E \sin(\beta/2),\beta).
\end{equation}

The explicit formula $(\ref{snr2})$ makes it easy to determine how to
optimally orient the detectors at a fixed value of $\beta$, and also
how the SNR at the optimal orientation varies with $\beta$.  The SNR
$(\ref{snr2})$ is maximized at $| \cos( 4 \delta)| = | \cos(2 \Delta)|
= 1$, and the optimum orientations are just configurations I and II
again, applying when $B < 0$ and $B > 0$ respectively.

\section{IMPLICATIONS FOR LIGO/VIRGO}
\label{implications}

The planned LIGO detectors in Hanford, Washington and Livingston,
Louisiana will have $\beta = 27.2^\circ$, $\delta = 44.9^\circ$ and
$\Delta = 28.2^\circ$ (the last two mod $90^\circ$) \cite{ligo_team}.
A graph of $\gamma(f)$ for this configuration is shown in
Fig.~\ref{ligo_gamma}.  In Fig.~\ref{ligo_thetas} we show what the
functions $\Theta_1$ and $\Theta_2$ look like at the value $\beta =
27.2^\circ$ appropriate for LIGO.  For any possible LIGO orientations,
$\gamma(f)$ will be a linear combination of these two functions as in
Eq.~$(\ref{gamma_general})$.  From Fig.~\ref{ligo_thetas} we can see
that there will be a zero of $\gamma(f)$ at $f \sim 70 \, {\rm Hz}$,
whose position is relatively insensitive to the orientations chosen.

The noise power spectral density in the LIGO detectors after some
years of operation might be roughly
\begin{equation}
\label{v_noise}
S_n(f) = \max \left[ S_m (f / f_m)^{-4}, S_m (f / f_m)^2 \right],
\end{equation}
where $S_m = 10^{-48} \, {\rm Hz}^{-1}$ and $f_m = 70 \, {\rm
Hz}$.  For frequencies less than $\sim 10 \, {\rm Hz}$ the noise will
be effectively infinite.  Eq.~$(\ref{v_noise})$ is a crude analytic
fit to the noise curve estimate for the ``advanced detectors'' given
by the LIGO team in Ref.~\cite{ligo_science}.  A more detailed model
of the noise is unnecessary because of the uncertainty as to
what the actual noise levels might be.

We now estimate what the effective site noise level $S_n^{\rm (eff)}$
[cf.~Eq.~$(\ref{site_noise})$] might be if - after an upgrade of the
LIGO facility - there are eventually two or three interferometers at
each site, as is planned.  The noise spectral density
$S_n(f)$ in each detector will be a sum of the
form\cite{ligo_science,300years}
\begin{equation}
S_n = S_{\rm seismic} + S_{\rm thermal} + S_{\rm shot} + S_{\rm gas} +
\ldots,
\end{equation}
including contributions due to seismic noise, thermal noise, photon
shot noise and residual gas noise, among others.  By contrast, the
typical value $S_n^\prime$ of the off-diagonal elements of the
spectral density matrix will be approximately a sum of terms that are
due to sources of noise which are strongly correlated between
different interferometers at the same site:
\begin{equation}
\label{correl}
S_n^\prime \alt S_{\rm seismic} + S_{\rm gas}.
\end{equation}
Since $S_{\rm seismic} + S_{\rm gas} \ll S_n$ when $f \agt 10 \, {\rm
Hz}$ for the advanced detectors of Ref.~\cite{ligo_science}, we find
from Eq.~(\ref{site_noise0}) that to a good approximation for LIGO,
\begin{equation}
\label{ligo_site_noise}
S_n^{\rm (eff)}(f)^{-1} \approx \sum_a S_n(f)_{aa}^{-1}.
\end{equation}
Here the sum is over the different detectors, and $a=1$ corresponds to
the primary, broadband interferometer with noise curve (\ref{v_noise}).
Now the second and/or third detectors at each site will most likely be
specialized ones such as dual-recycled detectors\cite{meers}, which
have high sensitivity only in some narrow frequency band.  Thus in the
relevant frequency band of $20 \, {\rm Hz} \alt f \alt 70 \, {\rm Hz}$
[cf.~Fig.~\ref{lastplot} below], it is likely that $S_n^{\rm
(eff)}(f) \approx S_n(f)_{11}$.  If, instead, three identical
broadband detectors are operated at each site (which is unlikely
initially), then $S_n^{\rm (eff)}(f) \approx S_n(f)_{11} / 3$, and the
$90\%$ confidence upper limit estimated below should be divided by $3$.

If we now make the plausible assumption that the SB has approximately
constant $\Omega_g(f)$ over the above waveband, then by
inserting Eq.~$(\ref{v_noise})$ into Eq.~$(\ref{snr1})$ we can
calculate the $90\%$ confidence upper limit that we can place on
$\Omega_g$.  Using the planned LIGO positions and orientations we thus
obtain
\FL
\begin{eqnarray}
\label{expected}
{}^{\rm(B)}\Omega_g^{(90\%)} & = & 1.65 \, { 5 \pi \over 4 \rho_c }
\left[ 2 {\hat \tau} \int_0^\infty df \, {\gamma(f)^2 \over f^6
S_n^{\rm (eff)}(f)^2 } \right]^{-1/2} \nonumber \\
& = & 5.1 \times 10^{-10} \, h_{75}^{-2} \, N
\left( {{\hat \tau} \over 10^7 \, {\rm s}} \right)^{-1/2},
\end{eqnarray}
where $h_{75}$ is the Hubble constant scaled to the value $75 \, {\rm
km\,s^{-1} Mpc^{-1}}$, $\hat \tau$ is the observation time, the
prefix ${\rm (B)}$ indicates broadband, and $N = S_m / 10^{-48} \,
{\rm Hz}^{-1}$.  Note that this is the upper bound obtained from
intersite cross-correlations, since as argued in Sec.~\ref{cnoise}, it
will not help much to include the information from intrasite
correlations.

This bound (\ref{expected}) is a reasonably conservative estimate of
what the LIGO sensitivity might be when the advanced detectors are
operating.  However the ultimate noise levels are largely unknown.
This is because the experimental techniques and technology, which are
presently somewhat far from the advanced level, will be steadily
improving.  To obtain an upper limit to ${}^{\rm (B)}
\Omega_g^{(90\%)}$ we take $S_n^{\rm (eff)}(f) \approx S_{\rm
gas}(f)$, which is a source of correlated noise that cannot be changed
once the beam tube is built.  Using $S_{\rm gas} \approx 2.2 \times
10^{-50} \, {\rm Hz}^{-1}$ \cite{stan} gives as a rather firm upper
limit for the LIGO sensitivity
\begin{equation}
\label{ultimate}
{}^{\rm (B)} \Omega_g^{\rm (90\%)} \ge 2 \times 10^{-13} N h_{75}^{-2}
\left( {{\hat \tau} \over 10^7 \, {\rm s}} \right)^{-1/2}.
\end{equation}

We now consider how ${}^{\rm(B)}\Omega_g^{(90\%)}$ would change if we
were to vary the angles $\beta$, $\delta$ and $\Delta$.  The integral
$(\ref{diagnostic})$ is positive for $\beta = 27.2^\circ$ with
the noise power spectral density $(\ref{v_noise})$.  Hence the optimal
configuration of the detectors is configuration II, i.e.~at $\delta =
\Delta = 0^\circ$ or $45^\circ$.  However LIGO already has $\delta
\approx 45^\circ$ (corresponding to parallel detectors because $\sigma_1 -
\sigma_2 = 2 \delta \approx 90^\circ$).  Also, as we have discussed in
Sec.~\ref{gamma_sec}, the dependence of $\gamma(f)$ on $\Delta$ is
very weak for $f \alt 100 \, {\rm Hz}$ and at $\beta = 27^\circ$,
see Eq.~$(\ref{gamma_general})$ and Fig.~\ref{ligo_thetas}.  Because
of these two facts the expected LIGO sensitivity at its planned
configuration of $\delta = 44.9^\circ$ and $\Delta = 28.2^\circ$ is
only $\sim 3\%$ less than the optimal sensitivity which would be
attained at $\delta = \Delta = 45^\circ$.  This follows from
Eq.~$(\ref{snr2})$ where we find that $B/A = 0.045 $ and $C/A = 7.8
\times 10^{-3}$.

The European VIRGO and GEO detectors, however, are currently planned
to have $2 \delta = \sigma_1 - \sigma_2 \approx 45^\circ$.  Such a
configuration will allow VIRGO/GEO to extract information from both
polarization components of burst and periodic gravitational waves, but
will severely reduce the sensitivity to the SB.  For these detectors
$\beta = 9.75^\circ$, and we find from Eq.~$(\ref{diagnostic})$ that
$B/A = 4.0 \times 10^{-4}$ and $C/A = 1.7 \times 10^{-6}$.  In
Fig.~\ref{virgo-geo} we plot the broadband signal to noise ratio
$(\ref{snr2})$ as a function of $\delta$ (assuming that $\Delta$ is
optimally chosen); it can be seen that at $\delta = 45^\circ$, the
sensitivity to the SB is reduced by about three orders of magnitude.
An intermediate choice of $\delta \sim 40^\circ$ would give a
broadband sensitivity to the SB comparable to that of LIGO, while
still allowing the detectors to access two polarization components
of the gravitational-wave field that are largely independent.

The total rotation angle $\Delta$ is unimportant for the broadband
sensitivity to the SB in the case of detectors which are sufficiently
close to each other [but {\it not} unimportant for the narrow-band
sensitivity, cf.~Eq.~(\ref{loose}) above].  However, for detectors
which are on separate continents, $\Delta$ will become important.  For
example in Fig.~\ref{long_baseline} we show the functions
$\Theta_1(f)$ and $\Theta_2(f)$ for $\beta = 79.5^\circ$, the value
appropriate for one of the two LIGO/VIRGO cross correlations.  Since
$\Theta_2$ is typically large compared to $\Theta_1$ in this case,
varying $\Delta$ will have a large effect on the sensitivity.  This is
confirmed by Eq.~$(\ref{snr2})$, where we find $B/A = -0.22$ and $C/A
= 18.3$.

Next we consider how the broadband sensitivity to the SB varies with
the angle $\beta$, assuming that the orientations are optimally
chosen.  Recall [cf.~Sec.~\ref{optimization}] that this means either
configuration I or II is chosen.  In Fig.~\ref{beta} we plot the SNR
$(\ref{snr2})$ as a function of $\beta$, for both configurations I and
II, again assuming the noise curve $(\ref{v_noise})$.  It can be seen
that the sensitivity falls off rapidly when the detectors become more
that a few thousand kilometers apart.  The sensitivity for the LIGO
separation is already a factor of $\sim 4$ worse than that for
coincident detectors.  Hence if there is ultimately a worldwide
network of detectors in America, Europe, Japan and Australia, only the
cross-correlations between proximate detectors like LIGO/LIGO or
VIRGO/GEO will be important.

The various conclusions that we have drawn are relatively insensitive
to the the detailed properties of the detector noise curve
$(\ref{v_noise})$ that we have assumed (except for the value of the
minimum frequency).  This is because the integral $(\ref{snr})$ is
dominated by contributions in the narrow frequency band $20 \, {\rm
Hz} \alt f \alt 70 \, {\rm Hz}$; see Fig.~\ref{lastplot}.  At these
frequencies thermal noise dominates over photon shot
noise\cite{ligo_science}, so using sophisticated optical
configurations of the interferometer such as dual
recycling\cite{meers} to reduce the shot noise will not help much.  It
thus will be important for the detectors to have good low frequency
sensitivity for the purposes of placing upper limits on the SB.

\section{conclusion}
\label{conclusion}

To place upper bounds on the strength of the gravitational stochastic
background, or perhaps to detect it, is but one of the aims of LIGO
and other detector systems.  Moreover, this background may well be very
weak compared to waves from astrophysical sources; consider by analogy
the weakness of the electromagnetic stochastic background in the
visible region of the spectrum.  However, these waves if detected
would be amongst the most interesting that the detectors would see.

The sensitivity to these waves is therefore just one of various
factors that need to be taken into account when choosing the fixed
orientations of interferometric detectors.  Nevertheless, in order to
be able to make a wise choice, it is important to know the effect for
the sensitivity of choosing this or that orientation.  One of the main
results of this paper is the simple expression (\ref{general_result})
describing this dependence.  In particular, for frequencies such that
the phase lag between the detectors is $\alt 1$, the sensitivity is
determined solely by the overlap of the polarization tensors of the
two detectors.  At higher frequencies the direction of the vector
which joins the detectors also becomes important.

We also determine the orientation of a pair of interferometers that
will optimize the sensitivity of the pair to the SB, and show that the
orientations which have been chosen for the LIGO detectors are close
to optimal.

\nonum
\section{ACKNOWLEDGMENTS}

The author is grateful to Kip Thorne for an introduction to the issue
of detector orientations, for many helpful discussions and comments on
the presentation, and also for the foresight of guessing, roughly,
what the optimal filtering method and implications for LIGO/VIRGO
might be, which to a large extent motivated and guided this papers
analysis.  Thanks are also due to Eric Poisson for some helpful and
detailed comments.  This work was supported in part by NSF grant
PHY-9213508.

\appendix{STATISTICAL FOUNDATIONS}
\label{stats}

\subsection{Overview}

In this Appendix we justify the method of optimal filtering described
in Sec.~\ref{processing} and the resulting signal to noise ratio
$(\ref{snr1})$.  We do this by deriving the probability distribution
for the spectrum $\Omega_g$ given the detectors' output ${\bf h}(t)$,
$p[\Omega_g | {\bf h}(t)]$.

First we transform from a continuous to a discrete description of the
measured data.  In the body of this paper we have treated the output
of the detector network as a continuous vector random process ${\bf
h}(t)$.  However, a real, discrete measurement will be of a finite
duration ${\hat \tau}$, and will contain frequencies only up to some
maximum frequency determined by the time resolution of the sampling.
In other words the output of the detectors will consist of the numbers
$h_{aj} = h_a(t_j = t_{\rm start} + j \Delta)$ for $1 \le a \le n_d$
and $0 \le j \le N$, where $n_d$ is the number of detectors, $t_{\rm
start}$ is the starting time, and $\Delta$ is the time resolution, of
the order of $10^{-4} \, {\rm s}$.  The number of samples per
interferometer is $N = {\hat \tau}/\Delta$.  We denote by ${\bf X}$
the vector of numbers $h_{aj}$, which we assume to have a multivariate
gaussian distribution with some variance-covariance matrix ${\bf
\Sigma}^\prime$.  This matrix is essentially the correlation matrix
$(\ref{c_h_def})$:
\begin{equation}
\label{sigmap}
\Sigma_{ai\,bj}^\prime = C_h(t_i - t_j)_{ab}.
\end{equation}
However if we take a finite Fourier transform, which amounts to making
a change of basis in the space of vectors ${\bf X}$, then ${\bf
\Sigma}^\prime$ corresponds instead to the spectral density matrix
$(\ref{s_h_def})$.

Now from Eq.~$(\ref{sn_sum})$, we have ${\bf \Sigma}^\prime = {\bf
\Sigma}_n^\prime + {\bf \Sigma}_s^\prime({\bf \Omega})$, where the
contribution ${\bf \Sigma}_s^\prime$ from the SB depends on the
spectrum $\Omega_g(f)$ as in Eq.~$(\ref{correlation})$.  [Because of
the finite frequency resolution of order $\sim 1/{\hat \tau}$, we
represent the function $\Omega_g(f)$ as a finite vector ${\bf
\Omega}$.]  What we want to do is to extract information about ${\bf
\Omega}$ from a measurement of ${\bf X}$.

There are two general approaches to the task of quantifying the
information obtained, in situations of this sort.  First, one can in
principle compute the probability distribution function (pdf) for
${\bf \Omega}$ given the measurement ${\bf X}$, $p({\bf \Omega} | {\bf
X})$.  This pdf then contains complete information about the
measurement.  However, the calculation of the pdf $p({\bf \Omega} |
{\bf X})$ is frequently difficult in practice; and so one has to
resort to instead calculating estimators (statistics) ${\hat {\bf
\Omega}}({\bf X})$ which are functions of ${\bf X}$, chosen so that
the pdf for their values given some value of ${\bf \Omega}$, $p({\hat
{\bf \Omega}} | {\bf \Omega})$, is peaked near ${\hat {\bf \Omega}} =
{\bf \Omega}$.  There are standard criteria for choosing such
estimators, see for example Refs.~\cite{WZ,Helstrom}.

Now suppose that, instead of one measurement of ${\bf X}$, we have $n$
measurements ${\bf X}_1, \ldots ,{\bf X}_n$.  A standard statistical
result is that in the Cramer-Frechet-Rao limit of $n \to \infty$
\cite{Helstrom,CFR}, the two approaches discussed above yield the same
unique result.  More precisely, the pdf $p({\bf \Omega} | {\bf X}_j)$
becomes a gaussian centered at ${\hat {\bf \Omega}}_{\rm ml}({\bf
X}_j)$, where ${\hat {\bf \Omega}}_{\rm ml}$ is the so-called maximum
likelihood estimator of ${\bf \Omega}$.  The variance-covariance matrix
${\bf \Sigma}_\Omega$ of this probability distribution depends on the
${\bf X}_j$ only through ${\hat {\bf \Omega}}_{\rm ml}({\bf X}_j)$,
${\bf \Sigma}_\Omega = {\bf \Sigma}_\Omega({\hat {\bf \Omega}}_{\rm
ml})$.  Conversely, the pdf $p({\hat {\bf \Omega}}_{\rm ml} | {\bf
\Omega})$ of the estimator ${\hat {\bf \Omega}}_{\rm ml}$ given some
value of ${\bf \Omega}$, becomes a gaussian centered at ${\bf \Omega}$
with width ${\bf \Sigma}_\Omega({\bf \Omega})$.  Thus the two
probability distributions, which are conceptually very different
objects, become effectively the same, and all one really needs to
calculate is the variance-covariance matrix ${\bf \Sigma}_\Omega({\bf
\Omega})$.

However, this simplifying Cramer-Frechet-Rao limit does not apply in a
straightforward manner to the calculation of $p({\bf \Omega} | {\bf
X})$ for measurements of the SB; there are a number of subtleties and
differences from the usual situations discussed in
Refs.~\cite{Helstrom,CFR}.  First, we have only one measurement of
${\bf X}$ instead of a large number $n$ of measurements.
Nevertheless, something like a ``large number of measurements limit''
does apply, which we discuss farther below.  Second, one needs to
address the issue of distinguishing between the contributions to ${\bf
\Sigma}^\prime$ from the detector noise and from the SB, since both
are unknown a priori.  Third, the CFR limit only applies to the extent
that our {\it a priori} knowledge of the variables being measured is
unimportant.  However, our {\it a priori} knowledge about the correlated
detector noise in intrasite correlations {\it is} important in the
calculation of $p({\bf \Omega} | {\bf X})$, and consequently this pdf
is not approximately gaussian [cf.~Eqs.~(\ref{prob}) and (\ref{pians})
below].

To resolve these complications, we now derive from first principles an
approximate expression for $p({\bf \Omega} | {\bf X})$.  We also
identify the conditions under which the approximations we make are
valid, and show that they will all be satisfied in the LIGO/VIRGO
context.  We assume throughout that the effect of the SB on the
detectors is small compared to the detectors intrinsic noise, so that
from Eq.~(\ref{est}), $\Omega_g(f) \ll 10^{-6}$.

\subsection{Calculation of $p({\bf \Omega} | {\bf X})$}

We start by considering the {\it frequency resolution} of the
measurement of $\Omega_g(f)$.  In a finite measurement of length
${\hat \tau}$ there are roughly $N = {\hat \tau} / \Delta$ independent
frequencies $f_j$, and we get what amounts to one measurement of each
$\Omega_j = \Omega_g(f_j)$.  Now if ${\hat \tau}$ is doubled, we
double the number of measurements, but we also double the number of
variables measured.  Clearly, the Cramer-Frechet-Rao limit of repeated
measurements of the same variables cannot be attained in this context
without further assumptions.  Below we shall argue that in the
LIGO/VIRGO context, to a good approximation,
\begin{equation}
\label{assume}
{\bf C}_h(\tau) \approx 0 \,\,\,\, \mbox{ for } \tau  > \tau_c,
\end{equation}
for some correlation time $\tau_c \ll {\hat \tau}$ \cite{caveat1}.  In
this case the unknown quantities to be measured are ${\bf C}_h(j
\Delta)$ for $0 \le j \le n = \tau_c / \Delta$, or equivalently ${\bf
S}_h(f)$ at $n$ different frequencies; and the number of measurements
of each of these variables is $\sim N/n = {\hat \tau} / \tau_c \gg 1$.

In order to explain Eq.~(\ref{assume}), we fix a value of $\tau_c$,
and decompose ${\bf C}_h(\tau)$ into
\FL
\begin{equation}
{\bf C}_h(\tau) = {\bf C}_h(\tau) \Theta(\tau_c - |\tau|) + {\bf
C}_h(\tau) \Theta(|\tau| - \tau_c),
\end{equation}
where $\Theta$ is the step function.  Let ${\bf S}_h = {\bf S}_h^< +
{\bf S}_h^>$ be the corresponding decomposition in frequency space, so
that ${\bf S}_h^<(f)$ is ${\bf S}_h(f)$ averaged over frequency scales
of the order of $1/\tau_c$.  Then from Eq.~(\ref{sn_sum}) there will
be two contributions to ${\bf S}_h^>(f)$:
\begin{equation}
\label{twoterms}
{\bf S}_h^>(f) = {\bf S}_s^>(f) + {\bf S}_n^>(f).
\end{equation}
The first term here will be small if $\Omega_g(f)$ is smooth over
frequency scales $\sim 1 / \tau_c$ [cf.~Eq.~(\ref{correlation}) ],
which will be the case for currently conceived of SB sources when
$\tau_c = 100 \, {\rm s}$, for example.  The second term in
Eq.~(\ref{twoterms}) will also be small for this value of $\tau_c$,
except near isolated frequencies corresponding to high-Q resonances
\cite{Rabb}, the effect of which can be neglected.

We now calculate the pdf for ${\bf \Omega}$ using Bayes rule; by
virtue of the condition (\ref{assume}), something like the CFR limit
will apply.  Let ${\hat {\bf \Sigma}}^\prime = {\bf X} \otimes {\bf
X}$, which is the maximum likelihood estimator of ${\bf
\Sigma}^\prime$.  This is a $N n_d \times N n_d$ matrix, where $n_d$
is the number of detectors.  Similarly the matrices ${\bf
\Sigma}_n^\prime$ and ${\bf \Sigma}_s^\prime$ are of dimension $N
n_d$, and are given by formulae analogous to (\ref{sigmap}).  However
because of the condition (\ref{assume}) the independent variables in
${\bf \Sigma}^\prime$ can be formed into a smaller matrix ${\bf
\Sigma}$, which is given by
\begin{equation}
\Sigma_{ai\,bj} = C_h[ (i-j) \Delta ]_{ab}
\end{equation}
for $0 \le i,j \le n = \tau_c / \Delta$.  Note that since ${\bf C}_h(-
\tau) = {\bf C}_h(\tau)^{\rm T}$, the matrix ${\bf \Sigma}$ contains
only $n_d^2 n$ independent variables.  We define the matrices ${\bf
\Sigma}_n$, ${\bf \Sigma}_s$ and ${\hat {\bf \Sigma}}$ analogously in
terms of ${\bf C}_n(\tau)$, ${\bf C}_s(\tau)$ and the estimator
\begin{equation}
\label{c_h_estimator}
{\hat C}_h(\tau)_{ab} \equiv {1 \over {\hat \tau}}
\int_{t_{start}}^{t_{\rm start} + {\hat \tau}} dt \,\, h_a(t + \tau)
h_b(t),
\end{equation}
which is defined for $|\tau| < \tau_c$.  The estimator ${\hat {\bf
C}}_h$ is the quantity that will be measured; it is a sufficient
statistic for ${\bf \Omega}$, and moreover its discrete counterpart
${\hat {\bf \Sigma}}({\bf X})$ is the maximum likelihood estimator of
${\bf \Sigma}$.

The joint pdf for ${\bf \Sigma}_n$ and ${\bf \Sigma}_s$ given ${\bf
X}$ is
\FL
\begin{equation}
\label{e1}
p({\bf \Sigma}_n,{\bf \Sigma}_s | {\bf X}) \, \propto \,\,
p^{\rm (0)}_n({\bf \Sigma}_n) \,p^{\rm (0)}_s({\bf \Sigma}_s)
\exp{\left[- {1 \over 2} \Lambda^\prime({\bf \Sigma}^\prime) \right]},
\end{equation}
where
\begin{equation}
\label{likelihood}
\Lambda^\prime({\bf \Sigma}^\prime) =\ln \det {\bf \Sigma}^\prime +
{\hat {\bf \Sigma}}^\prime:{\bf \Sigma}^{\prime\,-1},
\end{equation}
and $p^{\rm (0)}_n$ and $p^{\rm (0)}_s$ are the pdfs
that represent our {\it a priori} knowledge.  From the distribution
(\ref{e1}) we obtain the pdf $p({\bf \Omega} | {\bf X})$ for ${\bf
\Omega}$ given ${\bf X}$ by integrating over ${\bf \Sigma}_n$:
\begin{equation}
\label{e2}
p({\bf \Omega} | {\bf X}) = {\cal N} \int d{\bf \Sigma}_n \,
p({\bf \Sigma}_n,{\bf \Sigma}_s({\bf \Omega}) | {\bf X}),
\end{equation}
where ${\cal N}$ is a normalization constant.

The above formulae involve primed, $N n_d \times N n_d$ matrices.  We
now express them in terms of the corresponding, unprimed, $n n_d
\times n n_d$ matrices, by using the fact that all the matrices will
be approximately diagonal in frequency space.  That is, they will be
approximately diagonal in the indices $i,j$ after a finite Fourier
transform change of basis, but not in the indices $a,b$.  For example,
on a suitable basis,
\begin{equation}
\label{diagonalf}
\left( {\Sigma_n} \right)_{aI,bJ} \approx \delta_{IJ} \, S_n(f_I)_{ab},
\end{equation}
where $f_I = I / \tau_c$ for $1 \le I \le n$.  It is straightforward
to verify using Eqs.~(\ref{assume}) and (\ref{c_h_estimator}) that
\cite{caveat2}
\begin{equation}
\label{reduce_lambda}
\Lambda^\prime( {\bf \Sigma}^\prime) \approx k \, \Lambda( {\bf \Sigma}),
\end{equation}
where $k = {\hat \tau} / \tau_c = N / n$ is the effective
number of measurements, and
\begin{equation}
\label{likelihood1}
\Lambda({\bf \Sigma}) =\ln \det {\bf \Sigma} + {\hat {\bf
\Sigma}} \,: {\bf \Sigma}^{-1}.
\end{equation}
For example, the second term in $\Lambda^\prime( {\bf \Sigma}^\prime
)$ can be written as
\begin{eqnarray}
\label{expand3}
{\hat {\bf \Sigma}}^\prime : {\bf \Sigma}^{\prime\,-1} & = & \left(
\Sigma^{\prime\,-1} \right)^{\alpha \beta} X_\alpha X_\beta \nonumber
\\
\mbox{} & = & 2 \int_{-\infty}^\infty  df \, \,{\tilde {\bf
h}}(f)^\dagger \cdot {\bf S}_h(f)^{-1} \cdot {\tilde {\bf h}}(f) ,
\nonumber \\ \mbox{} & = & {\hat \tau} \int_{-\infty}^\infty  df \,\,
{\rm tr} \left[ {\bf S}_h(f)^{-1} \cdot {\hat {\bf S}}_h(f) \right].
\end{eqnarray}
Here we have used Eq.~(\ref{p2}), switched into the time domain, used
Eq.~(\ref{c_h_estimator}) and switched back \cite{caveat2}.  The result is
just $k {\hat {\bf \Sigma}} : {\bf \Sigma}^{-1}$, and the second term
in Eq.~(\ref{likelihood}) transforms analogously.

Now Eqs.~(\ref{e1}), (\ref{e2}), (\ref{reduce_lambda}) and
(\ref{likelihood1}) together yield a formal expression for $p({\bf
\Omega} | {\bf X})$.  To proceed further we need to invoke some
approximations.  The first of these is the quadratic approximation: we
expand the likelihood function $(\ref{likelihood1})$ to give
\FL
\begin{eqnarray}
\label{quad}
\Lambda({\bf \Sigma}) &=& \ln \det {\hat {\bf \Sigma}} + n \, n_d
\nonumber \\
\mbox{} & & + {1 \over 2} {\rm tr} \left[ {\hat {\bf \Sigma}}^{-1}
\cdot \delta {\bf \Sigma} \cdot {\hat {\bf \Sigma}}^{-1} \cdot \delta {\bf
\Sigma} \right] + {\cal O}(\delta {\bf \Sigma}^3),
\end{eqnarray}
where $\delta {\bf \Sigma} = {\bf \Sigma} - {\hat {\bf \Sigma}} = {\bf
\Sigma}_n + {\bf \Sigma}_s({\bf \Omega}) - {\hat {\bf \Sigma}}$.
Clearly this approximation will be good only for certain values of
${\bf \Sigma}_n$ and of ${\bf \Omega}$; we discuss below the
implications of this restriction.  First we derive the
conditions under which most of the probability of the pdf
\begin{equation}
\label{pdf2}
p({\bf \Sigma} | {\bf X}) \, \propto \, \exp \left[ - {k \over
2} \Lambda({\bf \Sigma}) \right]
\end{equation}
will be concentrated in the region ${\cal Q}$ where the quadratic
approximation (\ref{quad}) is valid, so that the normalization of the
pdf (\ref{pdf2}) can be calculated correctly using Eq.~(\ref{quad}).
If we let $\lambda_1, \ldots ,\lambda_n$ be the eigenvalues of ${\bf
\Sigma} \cdot {\hat {\bf \Sigma}}^{-1} - {\bf 1}$, then up to an
additive constant,
\begin{equation}
\Lambda({\bf \Sigma}) = - \sum_j \left[\ln (1 + \lambda_j) +
{1 \over 1 + \lambda_j} \right].
\end{equation}
 From this formula it can be seen that the quadratic regime ${\cal Q}$
is given by $\max_j |\lambda_j| < \varepsilon$ where $\varepsilon$ is
some small number.  If we demand that the total probability in ${\cal
Q}$ be $1 - \delta$ for some $\delta \ll 1$, and assume that the a
priori distribution $p^{\rm (0)}_n$ is not sharply peaked, then we
find the condition
\begin{equation}
\label{quad_condition}
N \agt {4 \over \varepsilon^2} \left( \log n + | \log \delta| \right).
\end{equation}
This condition will be just barely satisfied for, e.g, $\varepsilon =
\delta = 0.01$, ${\hat \tau} = 10^7 \, {\rm s}$, $\tau_c = 100 \, {\rm
s}$, and $\Delta = 10^{-4} \, {\rm s}$, which are values that are
appropriate for LIGO/VIRGO.  Now if ${\bf \Omega} \ll 10^{-6}$, then
the intgeral over ${\bf \Sigma}_n$ in Eq.~(\ref{e2}) will be dominated
by contributions from the region ${\cal Q}$, and so the approximation
is valid.  For ${\bf \Omega} \agt 10^{-6}$, the exact probability
$p({\bf \Omega} \, | \, {\bf X})$ is very small, and so the quadratic
approximation will still give a qualitatively correct result.  In
particular, the normalization of the pdf $p({\bf \Omega} \, | \, {\bf
X})$ resulting from Eq.~(\ref{quad}) will be approximately correct.

The next approximation involves consideration of the relative
magnitudes of various components of the measured autocorrelation
matrix ${\hat {\bf \Sigma}}$ [cf. Eq.~(\ref{c_h_estimator}) above].
We introduce the following notation: for any matrix ${\bf A}$, ${\bf
A}^\parallel$ denotes the matrix consisting of the diagonal subblocks
(in the indices $a,b$) of ${\bf A}$ corresponding to intrasite
correlations, and ${\bf A}^\perp = {\bf A} - {\bf A}^\parallel$
consists of the off-diagonal subblocks.  We also define ${\bf A}^D$ to
be the matrix of diagonal elements (in the indices $a,b$) of ${\bf A}$
and ${\bf A}^O = {\bf A}^\parallel - {\bf A}^D$.  Thus, the estimator
${\hat {\bf \Sigma}}$ decomposes into ${\hat {\bf \Sigma}}^D + {\hat
{\bf \Sigma}}^O + {\hat {\bf \Sigma}}^\perp$, where ${\hat {\bf
\Sigma}}^D$ contains the detector noises ${\hat {\bf
C}}_h(\tau)_{aa}$, ${\hat {\bf \Sigma}}^O$ contains the measured
intrasite correlations, and ${\hat {\bf \Sigma}}^\perp$ the measured
intersite correlations.  Now we expect the contribution of correlated
detector noise to ${\hat {\bf \Sigma}}^\perp$ to be very small,
cf.~the discussion in Sec.~\ref{extra} above.  Hence from
Eq.~(\ref{est}), ${\hat {\bf \Sigma}}^\perp \sim \, \varepsilon \,
{\hat {\bf \Sigma}}^D$, where
\begin{equation}
\label{small_1}
\varepsilon \approx \, {\Omega_g \over 10^{-6} }.
\end{equation}
For example, if the SB is just barely detectable, then $\varepsilon
\sim 10^{-4}$.  We similarly define in order of magnitude the small
parameter ${\tilde \varepsilon}$ by
\begin{equation}
\label{small_2}
{\hat {\bf \Sigma}}^O \sim \, {\tilde \varepsilon} \,{\hat {\bf
\Sigma}}^D,
\end{equation}
which we expect to be of the order of $S_{\rm gas}(f) / S_n(f) \sim
10^{-2}$ or smaller if $\Omega_g \alt 10^{-8}$; see
Eq.~(\ref{correl}).

In the expression (\ref{quad}) for the likelihood function, to leading
order in $\varepsilon$ we can replace the factors of ${\hat {\bf
\Sigma}}^{-1}$ by $({\hat {\bf \Sigma}}^\parallel)^{-1}$, so that the
cross terms between $\delta {\bf \Sigma}^\perp$ and $\delta {\bf
\Sigma}^\parallel$ vanish.  Note that to this order, $({\hat {\bf
\Sigma}}^{-1})^\parallel = ({\hat {\bf \Sigma}}^{\parallel})^{-1}$.
Hence, the pdf (\ref{e1}) splits into a product of two factors which
incorporate the measured intersite and the intrasite correlations.
The {\it a priori} pdf for the detector noise in Eq.~(\ref{e1}) can be
written as
\begin{equation}
\label{e3}
p^{\rm (0)}_n({\bf \Sigma}_n) \approx \,{\bf \delta}({\bf
\Sigma}_n^\perp) \,p^{\rm (0)}_{n,\parallel}({\bf \Sigma}_n^\parallel),
\end{equation}
since we expect ${\bf \Sigma}_n^\perp \approx 0$.  Using
Eqs.~(\ref{e1}), (\ref{e2}), (\ref{reduce_lambda}), (\ref{quad}) and
(\ref{e3}) we obtain
\FL
\begin{eqnarray}
\label{prob}
p(&{\bf \Omega}& | {\bf X})  \approx  \, {\cal N}_1 \,
p^{\rm (0)}_s({\bf \Omega}) \, p_C({\bf \Omega}) \nonumber \\
& & \times \exp{ {\rm tr} \left\{ -{k \over 4} \left[ ({\hat {\bf
\Sigma}}^\parallel)^{-1} \cdot \left( {\bf \Sigma}_s({\bf
\Omega})^\perp - {\hat {\bf \Sigma}}^\perp \right) \right]^2 \right\}
},
\end{eqnarray}
where ${\cal N}_1$ is a normalization constant.  Here
\FL
\begin{eqnarray}
\label{intrasite}
p_C({\bf \Omega}) & \equiv & \int d{\bf \Sigma}_n^\parallel \,\,
p^{\rm (0)}_{n,\parallel}({\bf \Sigma}_n^\parallel) \nonumber \\
& & \times \exp{ {\rm tr} \left\{ -{k \over 4} \left[ ({\hat {\bf
\Sigma}}^\parallel)^{-1} \cdot \delta {\bf \Sigma}^\parallel \right]^2
\right\}},
\end{eqnarray}
is a function representing the information from intrasite
correlations, and $\delta {\bf \Sigma}^\parallel = {\bf
\Sigma}_n^\parallel + {\bf \Sigma}_s({\bf \Omega})^\parallel - {\hat
{\bf \Sigma}}^\parallel$.

Now if ${\tilde \varepsilon} \gg \varepsilon$ as estimated above, then
most of the information we obtain from the SB will come from the
intersite correlation measurements.  In this case the factor $p_C({\bf
\Omega})$ in Eq.~(\ref{prob}) can be approximated to be constant [see
Eq.~(\ref{pians}) below].  We discuss the implications of the
resulting pdf for data-processing in subsection \ref{implic}.  However
it may turn out that ${\tilde \varepsilon} \alt \varepsilon$, which
from Eqs.~(\ref{small_1}) and (\ref{small_2}) will be the case if
$\Omega_g \agt 10^{-8}$.  Even if $\Omega_g \ll 10^{-8}$, it may
happen that the correlation coefficient between different detectors
for residual gas noise in the beam tube will be small compared to
unity, so that in the notation of Eq.~(\ref{correl}), $S_n^\prime \ll
S_{\rm gas}$ and again ${\tilde \varepsilon} \alt \epsilon$.
Alternatively, it may be possible to detect and take account of bursts
of outgassing from the beam tube walls.  More detailed discussions of
the possible magnitude of intrasite correlated noise can be found in
Ref.~\cite{paper}.  Because of the possibility that ${\tilde
\varepsilon} \alt \varepsilon$, we now derive an approximate
expression for $p_C({\bf \Omega})$.

\subsection{Information from intrasite correlations}

To evaluate the expression (\ref{intrasite}) we first make a change of
variables.  Let ${\hat {\bf {\cal C}}}$ be the matrix of correlation
coefficients
\begin{equation}
\label{corr_def1}
{\hat {\bf {\cal C}}} = ({\hat {\bf \Sigma}}^D)^{-1/2} \cdot {\hat
{\bf \Sigma}}^O \cdot ({\hat {\bf \Sigma}}^D)^{-1/2},
\end{equation}
and similarly define matrices ${\bf N}$ and ${\bf G}$ by replacing
${\hat {\bf \Sigma}}^O$ in Eq.~(\ref{corr_def1}) by ${\bf \Sigma}_n^D$
and ${\bf \Sigma}_s({\bf \Omega})^\parallel$ respectively.  Also let
\begin{equation}
\label{corr_def2}
{\bf {\cal C}} = ({\bf \Sigma}_n^D)^{-1/2} \cdot {\bf
\Sigma}_n^O \cdot ({\bf \Sigma}_n^D)^{-1/2}.
\end{equation}
We assume that the volume element in Eq.~(\ref{intrasite}) can be
written as
\FL
\begin{equation}
\label{volume_el}
p^{\rm (0)}_{n,\parallel}({\bf
\Sigma}_n^\parallel) \,\, d{\bf \Sigma}_n^\parallel \, \propto \,
p_{\rm noise}({\bf \Sigma}_n^D) \, p_{\rm corr}({\bf {\cal C}}) \, d
{\bf N} \, d {\bf {\cal C}}.
\end{equation}
To leading order in ${\tilde \varepsilon}$ the cross term between
$\delta {\bf \Sigma}^D$ and $\delta {\bf \Sigma}^O$ in the argument of
the exponential in Eq.~(\ref{intrasite}) vanishes, and so it becomes
proportional to
\FL
\begin{equation}
\label{arg1}
{\rm tr} \left[ \left( {\bf N} + {\bf G}^D - {\bf 1} \right)^2 \, +
\, \left( {\bf N}^{1/2} \cdot {\bf {\cal C}} \cdot {\bf N}^{1/2} +
{\bf G}^O - {\hat {\bf {\cal C}}} \right)^2 \right].
\end{equation}
Using the relations $k \gg 1$, ${\hat {\bf {\cal C}}} \sim {\tilde
\varepsilon} \ll 1$, ${\bf G} \sim \Omega_g / 10^{-6} \ll 1$, and
$|{\bf {\cal C}}| \alt 1$, and assuming that the pdf $p_{\rm
noise}({\bf \Sigma}_n^D)$ is slowly varying, we can approximately
carry out the integral over ${\bf N}$.  From Eqs.~(\ref{intrasite}),
(\ref{volume_el}) and (\ref{arg1}), the result is
\begin{eqnarray}
\label{intrasite2}
p_C({\bf \Omega}) & \approx & {\cal N}_2 \, \int d{\bf {\cal C}} \,\,
p_{\rm corr}({\bf {\cal C}}) \nonumber \\
& & \times \exp{ {\rm tr} \left\{ -{k \over 4} \left[ {\bf {\cal C}} +
{\bf G}^O - {\hat {\bf {\cal C}}} \right]^2 \right\}},
\end{eqnarray}
where ${\cal N}_2$ is a constant.  This expression is actually only a
good approximation when $|{\bf {\cal C}}| \ll 1$, but the integral
$(\ref{intrasite2})$ is dominated by values of ${\bf {\cal C}}$ close
to ${\hat {\bf {\cal C}}} - {\bf G}^O$ which is $ \ll 1$.

Now the behavior of the function (\ref{intrasite2}) depends strongly
on our {\it a priori} information about the correlation matrix ${\bf
{\cal C}}$.  As in Eq.~(\ref{diagonalf}), this matrix will be
approximately diagonal on a frequency basis:
\begin{equation}
{\cal C}_{aI,bJ} \approx \, \delta_{IJ} \, {\cal C}_{abI},
\end{equation}
where from Eq.~(\ref{corr_def2}),
\begin{equation}
{\cal C}_{abI} = { S_n(f_I)^O_{ab} \over \sqrt{ S_n(f_I)_{aa}
\, S_n(f_I)_{bb} } }.
\end{equation}
Since ${\bf S}_h(f)$ is positive definite, we have $| {\cal C}_{abI} | <
1$.  Moreover, the variables ${\cal C}_{abI}$ will be real whenever
\begin{equation}
\label{intra_assume1}
\langle n_a(t + \tau) \, n_b(t) \rangle \, = \, \langle n_a(t) \, n_b(t +
\tau) \rangle
\end{equation}
for all $t$ and $\tau$, and thus in particular they will be real if
there are no sources of noise that preferentially excite one detector
later than another one.  If we assume that such sources of noise are
negligible, then by inserting into Eq.~(\ref{intrasite2}) the pdf
\begin{equation}
p_{\rm corr}({\bf {\cal C}}) = \delta( {\rm Im} \,{\bf {\cal C}}) \,
p_{\rm corr}^\prime( {\rm Re} \,{\bf {\cal C}}),
\end{equation}
we obtain an equation again of the form (\ref{intrasite2}), but where
${\bf {\cal C}}$ and ${\hat {\bf {\cal C}}}$ are now understood to be
real.  [A factor of $\exp \left[ - k \, {\rm tr} \, ( {\rm Im} \,
{\hat {\bf {\cal C}}} )^2 /4 \right]$ is absorbed into ${\cal N}_2$.]

Next we assume that our {\it a priori} information about the
correlation coefficients ${\cal C}_{abI}$ is of the form ${\cal
C}_{\rm min} \le {\cal C}_{abI} \le {\cal C}_{\rm max}$, so that we
can take \cite{caveat3}
\FL
\begin{eqnarray}
\label{assumed_pdf}
p_{\rm corr}({\bf {\cal C}}) & & d {\bf {\cal C}}  =
\prod\limits_{I=1}^n \, \prod\limits_{a < b} \nonumber \\
\mbox{} \times & & \Theta( {\cal C}_{\rm max} - {\cal C}_{abI}) \,
\Theta( {\cal C}_{abI} - {\cal C}_{\rm min}) \, d {\cal C}_{abI},
\end{eqnarray}
where $\Theta$ is the step function.  Using Eqs.~(\ref{intrasite2})
and (\ref{assumed_pdf}) and the replacement ${\rm tr} \to
 2 \sum_{I=1}^n$ [cf.~Eq.~(\ref{expand3})], gives
\FL
\begin{eqnarray}
\label{pians}
& & p_C({\bf \Omega}) \propto \, \prod\limits_{I=1}^n \,
\prod\limits_{a < b} \nonumber \\
& & \times \left[ {\rm erf}\left( {\cal C}_{\rm max} \sqrt{k} -
\beta_{abI} \right) - {\rm erf}\left ({\cal C}_{\rm min} \sqrt{k} -
\beta_{abI} \right) \right],
\end{eqnarray}
where
\begin{equation}
\label{beta1}
\beta_{abI} = \sqrt{k} \left( {\hat {\cal C}}_{abI} - G^O_{abI} \right).
\end{equation}
If we define ${\hat {\bf S}}_h(f)$ to be twice the Fourier transform
of the estimator $(\ref{c_h_estimator})$ as in Eq.~$(\ref{s_h_def})$,
and put
\begin{equation}
\label{omega_est_intra}
{\hat \Omega}_{abI} = {5 \pi f^3 \over 4 \rho_c} \, {\hat
S}_h(f_I)^O_{ab},
\end{equation}
then we find from Eq.~(\ref{est}) that
\FL
\begin{eqnarray}
\label{beta2}
\beta_{abI} & = & {4 \rho_c \over 5 \pi f^3} \sqrt{k \over
{\hat S}_h(f_I)_{aa} {\hat S}_h(f_I)_{bb} }
 \times \left( {\hat \Omega}_{abI} - \Omega_I \right) \nonumber \\
& \approx & {\sqrt{k} \over 10^{-6}}
\left( {\hat \Omega}_{abI} - \Omega_I \right),
\end{eqnarray}
where $\Omega_I \equiv \Omega_g(f_I)$.

We now discuss the content of Eqs.~(\ref{pians}) -- (\ref{beta2}).  If
we make no assumption about the amount of correlated noise present,
and so take ${\cal C}_{\rm min} = -1$ and ${\cal C}_{\rm max} = 1$,
then we see using $k \approx 10^5$ that $p_C({\bf \Omega})$ is roughly
constant for $0 \le \Omega_I \alt 10^{-6}$.  Thus we obtain little
information.  More information about the correlated noise needs to be
input in order to constrain the strength of the SB.  One assumption
which may be valid is that
\begin{equation}
\label{intra_assume2}
S_n(f)_{ab} \ge 0,
\end{equation}
which can be enforced by setting ${\cal C}_{\rm min} = 0$ in
Eq.~(\ref{assumed_pdf}).  Equation (\ref{intra_assume2}) will hold if
there are no sources of noise ${\hat n}(t)$ which contribute an amount
$ + {\hat n}(t)$ to the output of one detector, and an amount $- {\hat
n}(t)$ to the output of another.  As example of such a source of noise
could be a mode of vibration of a suspension system that couples in a
suitable way the vibrations of mirrors in two different
interferometers.  It may be a valid assumption that all such sources
of noise will be negligible.  In this case we can use in
Eq.~(\ref{pians}) the values ${\cal C}_{\rm min} = 0$ and ${\cal
C}_{\rm max}=1$.  [The value chosen for ${\cal C}_{\rm max}$ is
unimportant as long as $c_{\rm max} \gg 1/\sqrt{k}$].  Then from
Eq.~(\ref{beta2}) we see that $\beta_{abI} \ll {\cal C}_{\rm max}
\sqrt{k}$ as ${\hat \Omega} \ll 10^{-6}$, and so to a good
approximation,
\begin{equation}
\label{pc1}
p_C({\bf \Omega})   \propto  \, \prod\limits_{I=1}^n \, \prod\limits_{a <
b} \left[ {1 \over 2} - {\rm erf}(- \beta_{abI}) \right].
\end{equation}

Essentially this pdf gives an upper bound on each $\Omega_I =
\Omega_g(f_I)$ of $\Omega_{I,{\rm max}} = \min_{a,b} \, {\hat
\Omega}_{abI}$, which is of the order of $\sim 10^{-6} {\tilde
\varepsilon}$ if $\Omega_g \alt 10^{-8}$.  As already mentioned, this
upper bound will be much worse than that obtained from intersite
correlations, unless the dimensionless correlation coefficient
${\tilde \varepsilon}$ is $\alt 10^{-4}$.  Moreover, the bound is only
obtained by making the specific assumptions (\ref{intra_assume1}) and
(\ref{intra_assume2}) about sources of correlated noise at one site.
For these reasons, in the following subsection on data-analysis we
consider only intersite correlations and take $p_C({\bf \Omega})
\approx \mbox{const}$.

We now show in more detail that very little information about the SB
is obtained if the assumptions (\ref{intra_assume1}) and
(\ref{intra_assume2}) are dropped.  As a simple model, we consider the
opposite extreme of assuming equal probability for all relative phases
in the contributions to the outputs of different detectors from
correlated sources of noise.  Thus, we take the {\it a priori} pdf to
be of the form \cite{caveat3}
\FL
\begin{eqnarray}
\label{assumed_pdf1}
p_{\rm corr}({\bf {\cal C}})& & d {\bf {\cal C}}= \prod\limits_{I=1}^n \,
\prod\limits_{a < b} \nonumber \\
\mbox{} \times & &
\exp \left[- {| {\cal C}_{abI}|^2 \over 2 {\cal C}_{\rm max}^2} \right]
 d ({\rm Re} \, {\cal C}_{abI}) \, d ({\rm Im} \, {\cal C}_{abI}),
\end{eqnarray}
where ${\cal C}_{\rm max}$ is the a priori maximum correlation.  Then from
Eq.~(\ref{intrasite2}) we obtain
\begin{equation}
\label{pc2}
p_C({\bf \Omega}) \, \propto \,  \prod\limits_{I=1}^n \,
\prod\limits_{a < b} \,\, \exp \left[ - {\left( {\rm Re} \, \beta_{abI}
\right)^2 \over 1 + k {\cal C}_{\rm max}^2 } \right].
\end{equation}
This function is of the form of Eq.~(\ref{simpler}) below, where, from
Eq.~(\ref{beta2}),
\begin{equation}
\sigma_I \approx 10^{-6} \sqrt{ 1 + k \, {\cal C}_{\rm max}^2 \over k}.
\end{equation}
This is roughly the minimum detectable value of $\Omega_g$ that can
be detected in a bandwidth of $\sim 1 / \tau_c$.  Thus, the function
(\ref{pc2}) is qualitatively similar to the pdf (\ref{simpler})
obtained from intersite correlations, with the simple change that the
minimum detectable value of $\Omega_g$ in any frequency band is
increased by a factor of $\sqrt{1 + k \, {\cal C}_{\rm max}^2}$.  If we take
${\cal C}_{\rm max}$ to be of order unity and so make no assumption about the
amount of correlated noise present, then the upper bounds on
$\Omega_g$ from intrasite correlations will be worse by a factor of
$\sqrt{k} \sim 300$ than those obtained from intersite correlations.
Only if ${\cal C}_{\rm max} \approx 1/\sqrt{k}$ will the two be comparable.
However, because of the possibility of weak, unknown sources of
noise, it is probably inappropriate to make such a strong assumption.

\subsection{Implications for data processing}
\label{implic}

The distribution (\ref{prob}) with $p_C = \mbox{const}$ is of the form
\FL
\begin{equation}
\label{simpler}
p(\Omega_I) = {\cal N}_3 \,p^{\rm (0)}_s(\Omega_I) \, \exp \left\{ -
\sum\limits_{I=1}^n
{(\Omega_I - {\hat \Omega}_I)^2 \over 2 \sigma_I^2} \right\},
\end{equation}
where $\Omega_I = \Omega_g(f_I = I / \tau_c)$, because the matrices
are all approximately diagonal in frequency.  Using
Eq.~$(\ref{correlation})$, we find that the argument of the exponential
in Eq.~(\ref{prob}) becomes
\begin{equation}
\label{cts}
- {{\hat \tau} \over 2} \int_0^\infty df \, {\rm tr} \left\{ \left[
{\hat {\bf S}}_h^{-1\, \parallel} \cdot \left( \Omega {\bar {\bf
\gamma}}^\perp - {\hat {\bf S}}_h^\perp \right) \right]^2 \right\},
\end{equation}
where ${\bar \gamma}_{ab} = 4 \rho_c \gamma_{ab} / (5 \pi f^3)$.
Hence we find, using $\int_0^\infty \, df \to (1 / \tau_c) \sum_I$,
that
\begin{equation}
\label{est1}
{1 \over \sigma_I^2} = k \, {\rm tr} \, \left[ \left(
{\hat {\bf S}}_h^{\parallel \, -1}(f_I) \cdot {\bar {\bf \gamma}}
\right)^2 \right],
\end{equation}
and
\FL
\begin{equation}
{\hat \Omega}_I = k \, \sigma_I^2 \, {\rm tr} \, \left[
{\hat {\bf S}}_h^{\parallel \, -1}(f_I) \cdot {\bar {\bf \gamma}} \nonumber
\cdot
{\hat {\bf S}}_h^{\parallel \, -1}(f_I) \cdot {\hat {\bf
S}}_h^\perp(f_I) \right].
\end{equation}

One would like to summarize the information contained in
Eq.~(\ref{simpler}) by calculating some kind of signal to noise ratio.
There are various, inequivalent ways of doing this.  For example, one
could calculate probability distributions for the quantities
$\Omega_{\rm max} = \max_I \Omega_I$ or ${\bar \Omega} = \sum_I
\Omega_I / n$; it is clear that the SNR for $\Omega_{\rm max}$ would
be worse than that for ${\bar \Omega}$.  What we shall in fact do is
calculate the probability distribution $p(\Omega | \Omega_I = \Omega)$
of $\Omega$ assuming that $\Omega_1 = \ldots = \Omega_n = \Omega$, as
this is the easiest to calculate.  Now if we ignore the {\it a priori}
information represented by $p^{\rm (0)}_s$ in Eq.~(\ref{simpler}) [so
that $p(\Omega_I)$ is a multivariate gaussian], then this is
equivalent to calculating the pdf for the weighted average
\begin{equation}
\label{mean2}
{\bar \Omega}_1 = \sum_I {\Omega_I \over \sigma_I^2} \left/ \sum_I {1
\over \sigma_I^2} \right.
\end{equation}
However when we include the information contained in $p^{\rm (0)}_s$,
the main effect is to truncate and renormalize\cite{explainSNR} the
pdfs for each ${\Omega_I}$, since all the $\Omega$'s must
be positive.  Hence a simple interpretation of the pdf $p(\Omega |
\Omega_I = \Omega)$ and the corresponding SNR in terms of the average
(\ref{mean2}) not possible.  Nevertheless we suspect that $p(\Omega |
\Omega_I = \Omega)$ approximately represents the probability
distribution of some type of average of $\Omega_g(f)$.

We now insert the assumption $\Omega_I = \Omega$ (constant) into
Eq.~(\ref{simpler}), which gives a pdf of the form
\begin{equation}
\label{simple1}
p(\Omega \, | \, \Omega_I = \Omega) \propto \, \Theta(\Omega)  \exp
\left[ {(\Omega - \Omega_M)^2 \over (2 \sigma^2) } \right].
\end{equation}
Here the quantity
\begin{equation}
\Omega_M = \sum_I {{\hat \Omega}_I \over \sigma_I^2} \left/ \sum_I {1
\over \sigma_I^2} \right.
\end{equation}
is the statistic that should be
calculated to estimate the value of $\Omega$.  We obtain
\begin{equation}
\Omega_M \, \propto \, \int_0^\infty df \, {\rm tr} \,\,\left[ {\hat
{\bf S}}_h^\perp
\cdot {\hat {\bf S}}_h^{-1 \, \parallel}  \cdot {\bar {\bf
\gamma}}^\perp \cdot {\hat {\bf S}}_h^{-1 \, \parallel} \right].
\end{equation}
This can be written in the time domain as
\begin{equation}
\label{omega_est}
\Omega_M \propto \, \int dt \, \int d\tau \, H_a(t + \tau)
L_{ab}(\tau) H_b(t),
\end{equation}
where ${\tilde {\bf L}}(f) = {\bf \gamma}(f)^\perp$, and
\begin{equation}
\label{Hfilter}
{\tilde {\bf H}}(f) = {1 \over f^{3/2}} {\hat {\bf
S}}_h^{-1}(f)^\parallel \cdot {\tilde {\bf h}}(f).
\end{equation}
The functions $L_{ab}(\tau)$ are the sliding delay functions discussed
in Sec.~\ref{processing}.  Note that $\Omega_M$ is constructed from
the measured correlations in the following way: the intrasite
correlations are used only to estimate the detector noise matrix ${\bf
S}_n^\parallel$, and then this noise matrix is used together with the
measured inter-site correlations to estimate $\Omega_M$.  However, the
matrix ${\bf S}_n^\parallel$ will typically be approximately diagonal,
and so neglecting the intrasite correlations will give only a
negligible error.

Finally the intersite SNR $\Omega_M / \sigma$ can be obtained from
Eq.~(\ref{cts}) \cite{explainSNR}.  If we assume that ${\hat {\bf
S}}_h(f)^\perp = \Omega_{\rm real}(f) {\bar {\bf \gamma}}(f)^\perp$,
so that all the intersite correlations we measure come from the SB,
then the SNR takes the form
\begin{eqnarray}
\label{snr3}
{S^2 \over N^2} & = & {\hat \tau} \left( 4 \rho_c \over 5 \pi \right)^2
\, \int_0^\infty df \, {\Omega_{\rm real}(f)^2 \over f^6} \nonumber \\
\mbox{} & & \times {\rm tr} \left[ \left( {\bf \gamma}(f)^\perp \cdot
{\hat {\bf S}}_h^{-1}(f)^\parallel \right)^2 \right].
\end{eqnarray}
Using the fact that the matrix $\gamma(f)_{ab}$ will be constant
on each subblock corresponding to 2 detector sites, one can
derive Eqs.~$(\ref{snr1})$ and $(\ref{site_noise0})$ from
Eq.~(\ref{snr3}).

\appendix{CALCULATION OF THE \\
OVERLAP REDUCTION FUNCTION}
\label{gamma_calc}

The $a$th detector is characterized by its position ${\bf x}_a$ and by
the tensor ${\bf d} = ({\bf u} \otimes {\bf u} - {\bf v} \otimes {\bf
v})/2$, where ${\bf u}$ and ${\bf v}$ are unit vectors in the
direction of its arms.  In terms of these quantities, the overlap
reduction function is given, from Eq.~$(\ref{gamma_def})$, by
\begin{eqnarray}
\gamma_{ab}(f) = {5 \over 8 \pi} \sum_A & & \int d^2 \Omega \,
({\bf d}_a : {\bf e}^{A,{\bf n}} )\,\, ({\bf d}_b : {\bf e}^{A,{\bf n}})
\nonumber \\
\mbox{} & & \times \exp \left[ 2 \pi i f {\bf n} \cdot ({\bf x}_a - {\bf
x}_b) \right].
\end{eqnarray}
If we write ${\bf x}_a - {\bf x}_b = d \, {\bf m}$ where ${\bf m}$ is a unit
vector, and put $\alpha = 2 \pi f d$ (in units in which $c = 1$), then
we obtain
\begin{equation}
\label{trick}
\gamma_{ab}(f) = d_{a\, ij} \,\, \Gamma_{ijkl}(\alpha,{\bf m}) \,\, d_{b \,
kl},
\end{equation}
where
\FL
\begin{equation}
\Gamma_{ijkl}(\alpha,{\bf m})  = {5 \over 8 \pi} \sum_A \int d^2
\Omega \, e^{A,{\bf n}}_{ij} e^{A,{\bf n}}_{kl} \, e^{i \alpha {\bf n}
\cdot {\bf m}}.
\end{equation}

This integral can be evaluated by the standard method of writing down
the most general possible answer:
\FL
\begin{eqnarray}
\label{Gamma_ans}
\Gamma_{ijkl}(& \alpha &,{\bf m}) = A(\alpha) \delta_{ij} \delta_{kl} +
B(\alpha) \left[ \delta_{ik} \delta_{jl} + \delta_{il} \delta_{jk}
\right] \nonumber \\
\mbox{} &+& C(\alpha) \left[ \delta_{ij} m_k m_l + \delta_{kl} m_i m_j
\right] + D(\alpha) m_i m_j m_k m_l \nonumber \\
\mbox{} &+& E(\alpha) \left[ \delta_{ik} m_j m_l + \dots + \delta_{jl}
m_i m_k \right].
\end{eqnarray}
One might expect to have to include a term proportional to $w_{il}
w_{jk} + w_{ik} w_{jl}$ where $w_{ij} = \varepsilon_{ijk} m^k$, but in
fact this tensor is a linear combination of the five tensors included
above.  Contracting Eq.~(\ref{Gamma_ans}) with the five different
tensorial expressions that appear on its right hand side yields a
system of linear equations for $A(\alpha) \ldots E(\alpha)$ which
involves scalar integrals that are straightforward to evaluate.
Solving these equations and substituting the results back into
Eqs.~(\ref{trick}) and (\ref{Gamma_ans}) gives
\begin{eqnarray}
\label{general_result}
\gamma_{ab}(f) &=& \rho_1(\alpha) {\bf d}_a : {\bf d}_b + \rho_2(\alpha)
{\bf m} \cdot {\bf d}_a \cdot {\bf d}_b \cdot {\bf m} \nonumber \\
\mbox{} & & + \rho_3(\alpha)  ( {\bf m} \cdot {\bf d}_a \cdot {\bf m} )
 ( {\bf m} \cdot {\bf d}_b \cdot {\bf m} ).
\end{eqnarray}
The functions $\rho_j(\alpha)$ are linear combinations of the
spherical bessel functions:
\FL
\begin{equation}
\rho_1(\alpha) = 5 j_0(\alpha) - 2 j_1(\alpha) / \alpha +
5 j_2(\alpha) / \alpha^2,
\end{equation}
\FL
\begin{equation}
\rho_2(\alpha) = - 10 j_0(\alpha) + 40 j_1(\alpha) / \alpha - 50
j_2(\alpha) / \alpha^2,
\end{equation}
and
\FL
\begin{equation}
\rho_3(\alpha) = 5 j_0(\alpha)/2 - 25 j_1(\alpha) / \alpha +
175 j_2(\alpha) / (2 \alpha^2).
\end{equation}

The result (\ref{general_result}) applies to any gravitational wave
antennas, such as interferometers with non-perpendicular arms and
arbitrary orientations, or resonant bar antennas.  The special case of
two resonant bar antennas, for which each ${\bf d}_a \propto {\bf 1} -
3 {\bf n}_a \otimes {\bf n}_a$ for some vector ${\bf n}_a$, has been
previously derived by Michelson \cite{mi}.  As can be seen from
Fig.~\ref{rho_fns}, the first term in Eq.~(\ref{general_result})
dominates for $\alpha \alt 1$ (unless ${\bf d}_a : {\bf d}_b \approx
0$).  A similar simplification applies for $\alpha \gg 1$: we have
\begin{equation}
\gamma_{ab}(f) = 5 j_0(\alpha) \,{\bf d}_a^\perp : {\bf d}_b^\perp \,+
{\cal O}(\alpha^{-2}),
\end{equation}
where ${\bf d}_a^\perp$ denotes the tracefree part of the projection
$(\delta_{ik} - m_i m_k) (\delta_{jl} - m_j m_l) d_{a\,kl}$ of ${\bf
d}_a$ perpendicular to ${\bf m}$.  The fact that $\gamma(f)$ does not
depend on the components of ${\bf d}_a$ and ${\bf d}_b$ parallel to
${\bf m}$ is due to the fact that the cross correlation at $\alpha \gg
1$ is dominated by modes whose wavevectors are nearly parallel to ${\bf m}$,
as discussed in Sec.~\ref{special_cases}.

To apply Eq.~(\ref{general_result}) to terrestrial detectors, we need
to choose a coordinate system and express the tensors ${\bf d}_1$,
${\bf d}_2$ and ${\bf m}$ in terms of the angles $\delta$, $\Delta$
and $\beta$ defined in Sec.~\ref{gamma_sec}.  A convenient choice is
to take the detectors to be located at $\theta = \pi/2 \pm \beta/2$
and $\phi = 0$, where $r,\theta,\phi$ are spherical polar coordinates
with origin at the earth's center, so that the unit vector in the
direction joining the detectors is ${\bf m} = {\bf e}_z$.  If we let
${\bf e}_{\hat r}$, ${\bf e}_{\hat \theta}$ and ${\bf e}_{\hat \phi}$
be the usual basis of orthonormal vectors, and define
\begin{eqnarray}
\label{d_def}
{\bf d}(\sigma,\theta,\phi) & = & \sin (2 \sigma) ( {\bf e}_{\hat \theta}
\otimes {\bf e}_{\hat \theta} - {\bf e}_{\hat \phi} \otimes {\bf
e}_{\hat \phi})/2 \nonumber \\
\mbox{} & & - \cos (2 \sigma) ( {\bf e}_{\hat \theta}
\otimes {\bf e}_{\hat \phi} + {\bf e}_{\hat \phi} \otimes {\bf
e}_{\hat \theta})/2,
\end{eqnarray}
then we can take ${\bf d}_1 = {\bf d}(\Delta + \delta, \pi/2 +
\beta/2,0)$ and ${\bf d}_2 = {\bf d}(\Delta - \delta, \pi/2 -
\beta/2,0)$.  Inserting these expressions into the
formula (\ref{general_result}) yields, after some manipulation, the
result $(\ref{gamma_general})$, where the functions $g_j(\alpha)$ are
\begin{equation}
\label{g1}
g_1(\alpha) = {5 \over 16} f(\alpha) \cdot (-9,-6,9,3,1),
\end{equation}
\begin{equation}
\label{g2}
g_2(\alpha) = {5 \over 16} f(\alpha) \cdot (45,6,-45,9,3),
\end{equation}
\begin{equation}
\label{g3}
g_3(\alpha) = {5 \over 4} f(\alpha) \cdot (15,-4,-15,9,-1),
\end{equation}
and
\FL
\begin{equation}
\label{fdef}
f(\alpha) = ( \alpha \cos \alpha, \alpha^3 \cos \alpha, \sin \alpha,
\alpha^2 \sin \alpha, \alpha^4 \sin \alpha) / \alpha^5.
\end{equation}

 From Eq.~$(\ref{gamma_general})$ it is straightforward to evaluate the
Fourier transform of the overlap reduction function, which gives the
sliding delay function discussed in Sec.~\ref{processing},
\begin{equation}
L_{ab}(\tau) = \int_{-\infty}^{\infty} df \, e^{-2 \pi i f \tau}
\gamma_{ab}(f).
\end{equation}
The function $L_{ab}(\tau)$ vanishes for $|\tau| > d = 2 r_E \sin
(\beta/2)$.  For $|\tau| < d$, $L_{ab}(\tau)$ is given by
Eqs.~$(\ref{gamma_general})$ to $(\ref{theta2})$, where the functions
$g_j$ are now
\begin{equation}
g_1(\tau) = {5 \over 32 d} ( 1 + 3 v + {3 \over 8} v^2),
\end{equation}
\begin{equation}
g_2(\tau) = {5 \over 32 d} ( 3 - 3 v - {15 \over 8} v^2),
\end{equation}
\begin{equation}
g_3(\tau) = {5 \over 32 d} ( -4 + 8 v - {5 \over 2} v^2),
\end{equation}
and $v = 1 - \tau^2 / d^2$.

\newpage
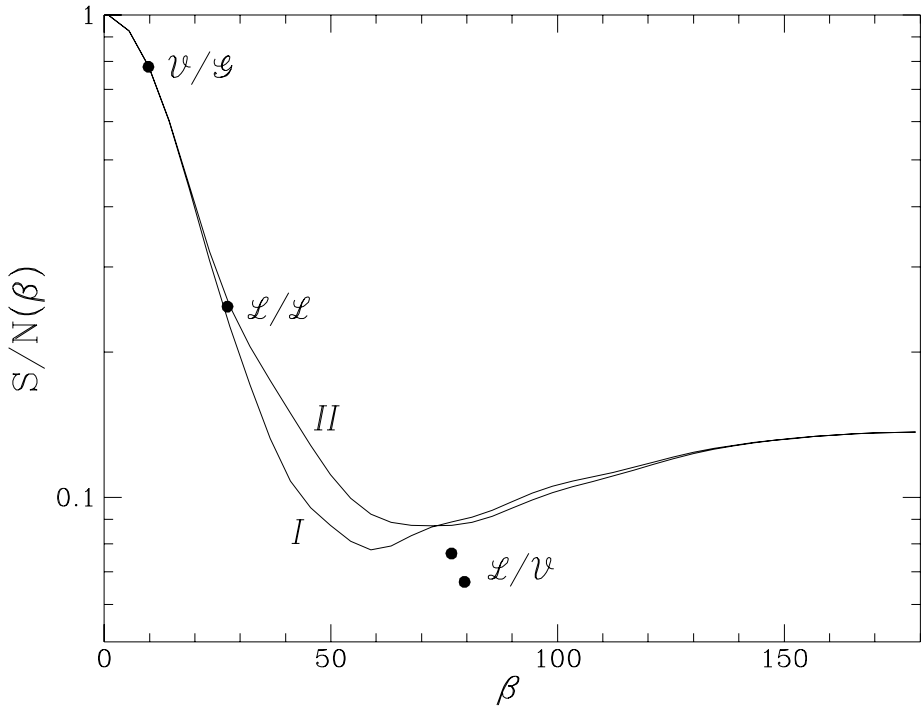
\figure{The broadband signal to noise ratio for a pair of detectors as
a function of the angle $\beta$ subtended between them at the center
of the earth, normalized to unity for coincident detectors.  Curve I
corresponds to each detector having an arm along the arc of the great
circle that joins them, and curve II corresponds to each one having an
arm at $45^\circ$ to this arc.  The optimal configuration is II for
$\beta \alt 70^\circ$ (except very close to $\beta = 0$), and I for
larger values of $\beta$.  The point ${\cal L}/{\cal L}$ and the two
points marked ${\cal L}/{\cal V}$ show the expected sensitivities of
the LIGO detector pair and of both LIGO/VIRGO detector pairs, with
their current orientations.  The point ${\cal V}/{\cal G}$ shows the
sensitivity the detector pair VIRGO/GEO would have if the orientations
were chosen optimally for the stochastic background (which will
probably not be the case, since optimization for the stochastic
background implies sensitivity to only one of the waves' two
polarizations, and a corresponding loss of information when studying
non-stochastic waves).
\label{beta}}

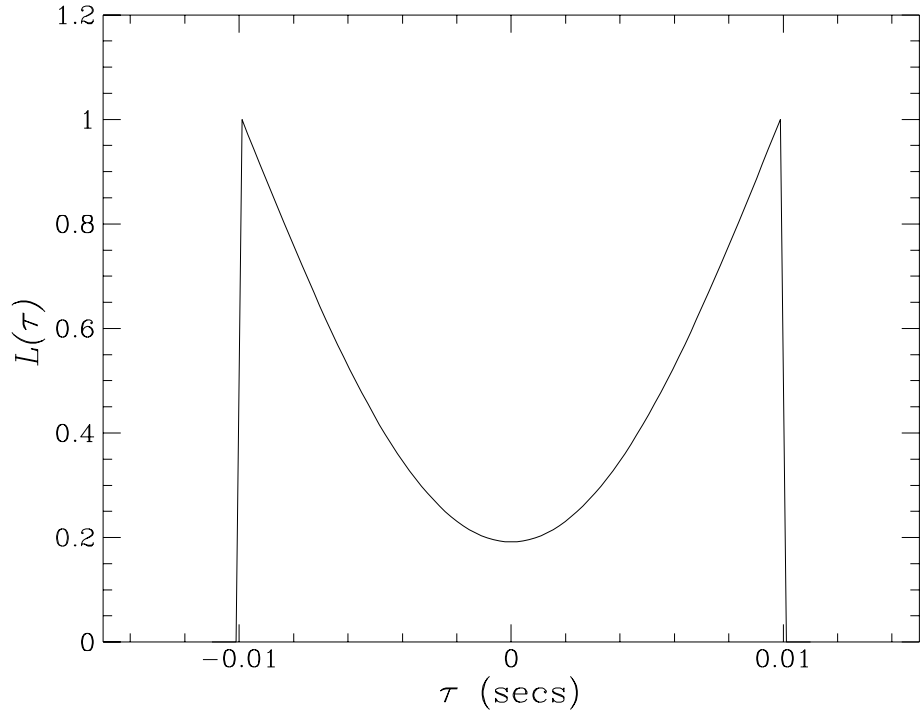
\figure{The ``sliding delay function'' $L(\tau)$ for the LIGO pair of
interferometers.  To maximize the broadband sensitivity to the
stochastic background, the cross correlation with a time delay $\tau$
between the detector pair must be integrated against this function;
see text.
\label{delay}}

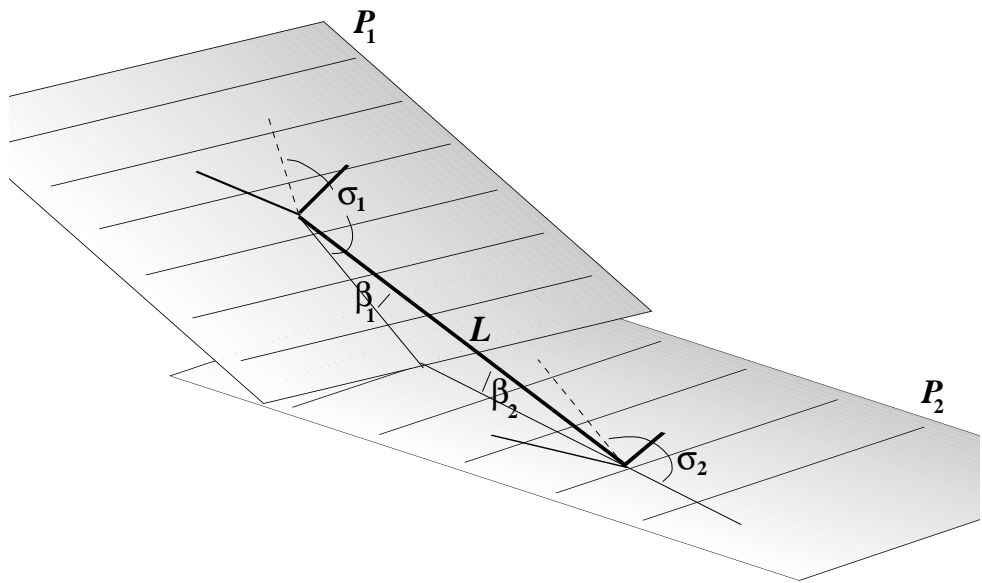
\figure{The angles $\sigma_1$, $\sigma_2$, $\beta_1$, and $\beta_2$
formed by a pair of interferometric detectors and the line $L$
which joins them.
\label{angles}}

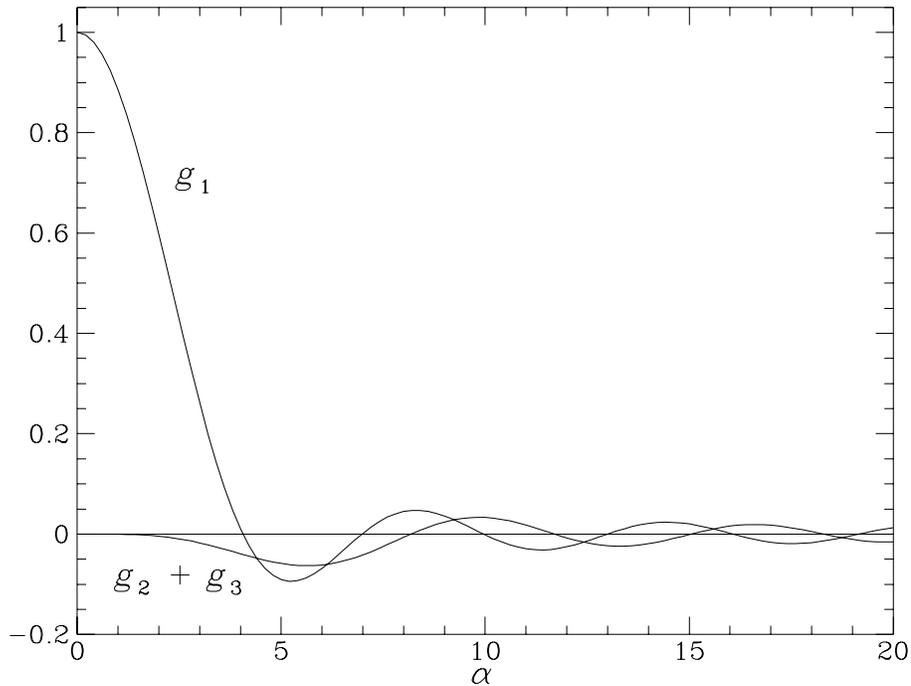
\figure{For coplanar detectors the overlap reduction function is of the
form $\gamma(f) = \cos(4 \delta) g_1(\alpha) + \cos(4 \Delta) \left[
g_2(\alpha) + g_3(\alpha) \right]$, where $\delta$ and $\Delta$ are
the relative and total rotation angles, and $\alpha$ is the phase lag
between the detectors at frequency $f$; see text.  Here we plot the
functions $g_1(\alpha)$ and $g_2(\alpha) + g_3(\alpha)$.
\label{planar}}

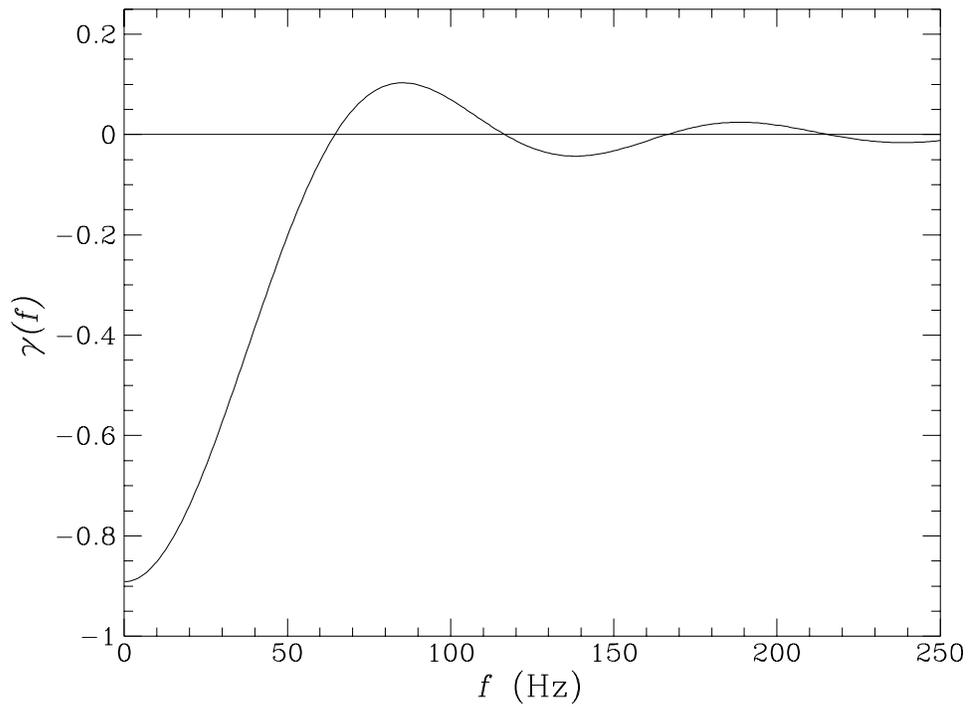
\figure{The overlap reduction function for the LIGO detectors using
their currently planned orientations.
\label{ligo_gamma}}

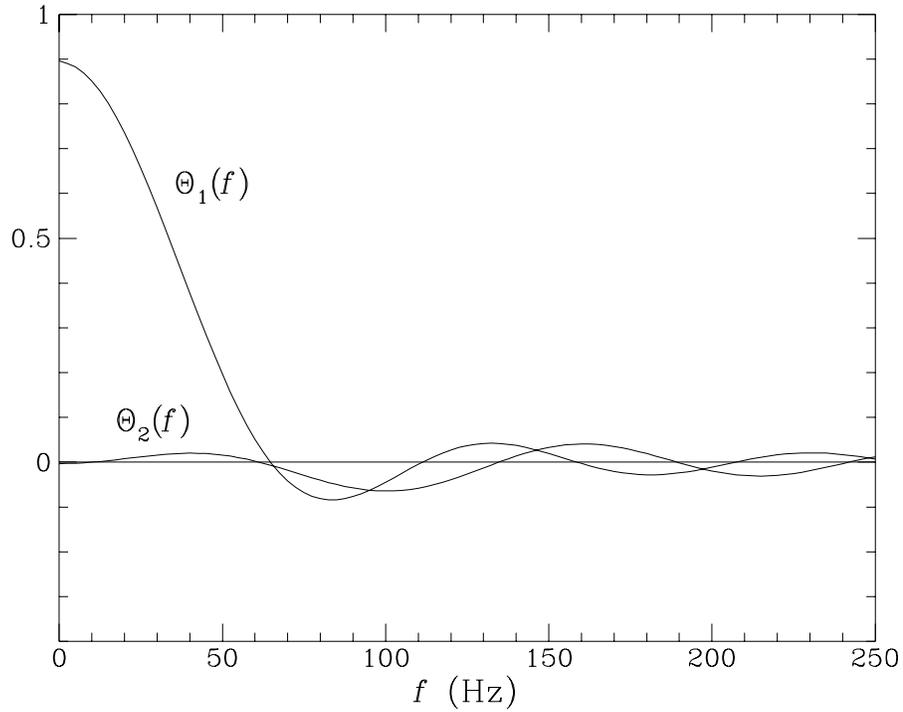
\figure{For {\it any} orientation angles $\delta$, $\Delta$ of the LIGO
detectors, the overlap reduction function will be given in terms of
two functions of frequency $\Theta_1(f)$ and $\Theta_2(f)$ by
$\gamma(f) = \cos (4 \delta) \Theta_1(f) + \cos(4 \Delta)
\Theta_2(f)$.  Here we show the functions $\Theta_1(f)$ and
$\Theta_2(f)$.
\label{ligo_thetas}}

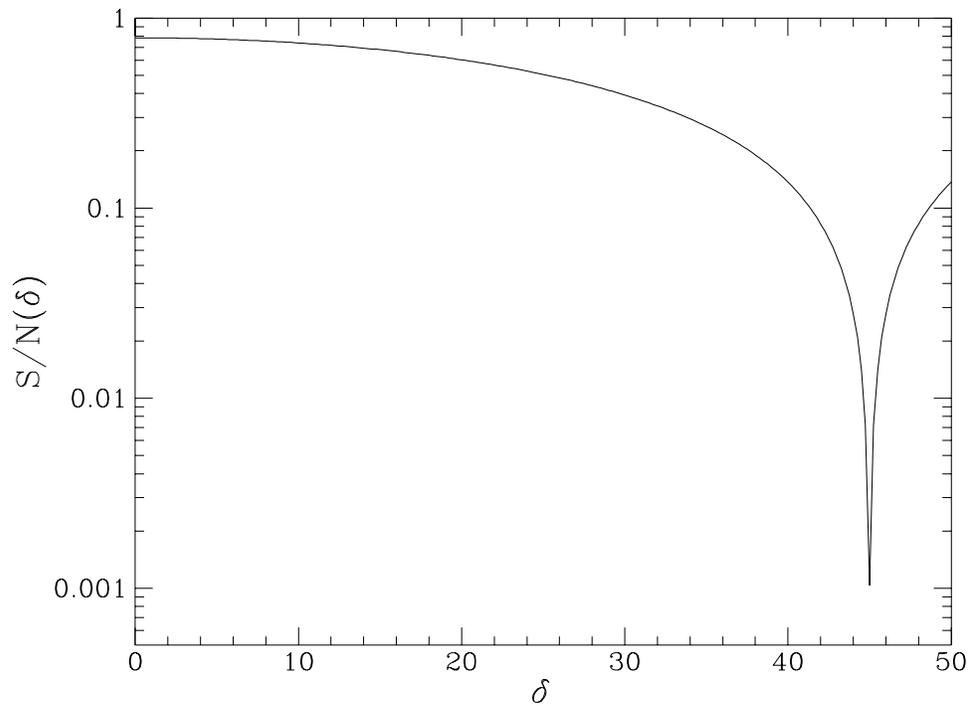
\figure{The broadband signal to noise ratio for the VIRGO/GEO detector
pair as a function of the relative rotation angle $\delta$, assuming
(i) that the total rotation angle $\Delta$ is optimally chosen, and
(ii) the advanced detector noise curve of Ref.~\cite{ligo_science}.
The normalization is to unity for coincident, aligned detectors.
\label{virgo-geo}}

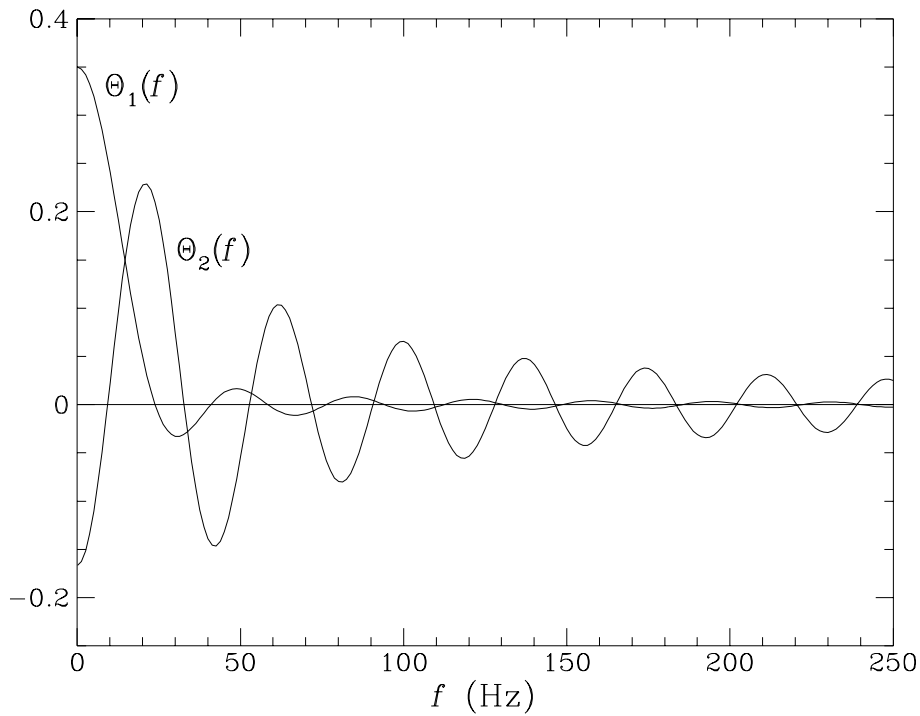
\figure{The functions $\Theta_1(f)$ and $\Theta_2(f)$ appropriate for
the LIGO detector in Hanford, Washington together with the VIRGO
detector in Pisa, Italy.  The fact that $\Theta_2(f)$ is large
compared to $\Theta_1(f)$ for most frequencies implies that the
sensitivity of the detector pair to the SB depends strongly in this
case on the total rotation angle $\Delta$ of the two detectors with
respect to the line joining them; see text.
\label{long_baseline}}

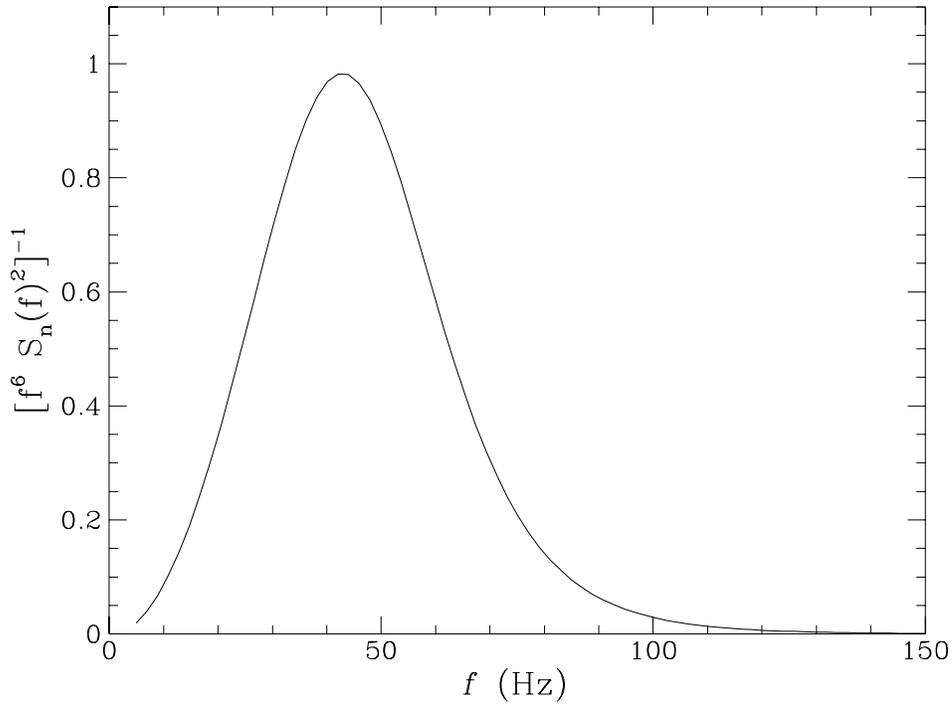
\figure{A plot of the quantity $f^{-6} S_n(f)^{-2}$ with arbitrary
normalization as a function of frequency.  It is this quantity that
must be integrated against the square of the overlap reduction
function to determine the broadband response of a detector pair to the
stochastic background, when one assumes a constant,
Harrison-Zel'dovich spectrum; see Eq.~(\ref{snr}).  It is clear that the
highest sensitivity is limited to a narrow band of frequency between
$\sim 20 \, {\rm Hz}$ and $\sim 70 \,{\rm Hz}$.
\label{lastplot}}

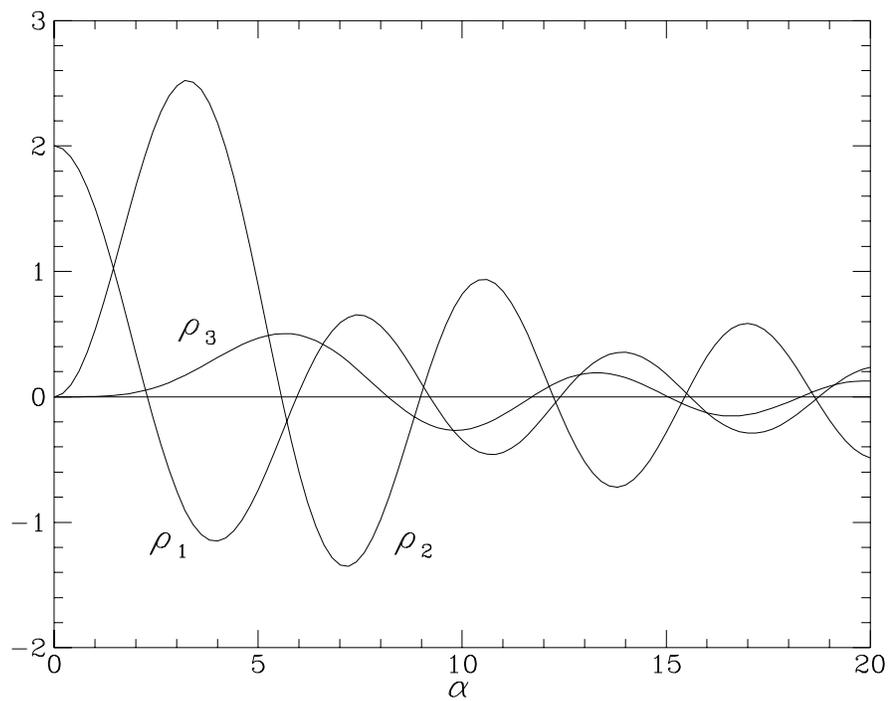
\figure{A graph of the functions $\rho_1(\alpha)$, $\rho_2(\alpha)$ and
$\rho_3(\alpha)$.
\label{rho_fns}}


\begin{references}
\bibitem{equiv_results} Eqs.~$(\ref{correlation})$ and $(\ref{gamma_def})$
are essentially equivalent to Eqs.~(4.5) and (4.6) of
Ref.~\cite{paper}, and can be obtained from those by (i) taking a
Fourier transform, (ii) dividing by the square of the transfer function
$|{\tilde B}(f)|$ to convert from correlations of light intensities to
correlations of strain amplitudes, (iii) replacing the cosine with a
complex exponential, and (iv) correcting for the difference in
normalization convention for $\gamma(f)$ between Ref.~\cite{paper} and
this paper.
\bibitem{disclaimer} It is not quite true that the contributions ${\bf
S}_n(f)$ and ${\bf S}_s(f)$ are indistinguishable.  The structure of
the matrix ${\bf S}_s(f)$ at a fixed frequency is tightly constrained
by Eq.~$(\ref{correlation})$, so the whole (off-diagonal) matrix can
be determined from just one element.  Moreover all the off-diagonal
elements must be real.  If we measure the off-diagonal part of ${\bf
S}_h(f)$ to be nonzero and it does not have the form
(\ref{correlation}), then we know that we are seeing noise that is not
due to the SB.  This fact will be helpful in identifying noise in
inter-site correlations (if there is any inter-site correlated noise).
It will not be as useful for identifying noise in intra-site
correlations, because in this case the constraint is just that all the
off-diagonal elements of ${\bf S}_A(f)$ be equal and positive, and
this is likely to be satisfied by some sources of correlated noise.
\bibitem{commenta} These designations must not be taken too literally
however, since for $\beta = 180^\circ$, the angle $\Delta$ is what
would normally be called the relative rotation; see Eq.~(\ref{d_def}).
\bibitem{caveat1} If the condition (\ref{assume}) were not valid, one
could still in principle integrate over $\Omega_g^>(f)$ to derive the
reduced pdf for $\Omega_g^<(f)$ [in the notation of
Eq.~(\ref{twoterms})].  However, in this case the CFR limit need not
apply.
\bibitem{caveat2}  There will be corrections due to aliasing which can
be made small by increasing $\tau_c$ by a factor of a few, so that
${\bf C}_h(\tau) \approx 0$ for $| \tau | > \tau_c / \mbox{(a few)}$.
\end{references}
\end{document}